\documentclass[numberedappendix]{emulateapj}
\usepackage{natbib,graphicx,times,amsmath}
\slugcomment{Submitted to the Astronomical Journal on October 12, 2005; Accepted February 17, 2006}

\def\arcdeg{\hbox{$^\circ$}}

\def\arcsec{\hbox{$^{\prime\prime}$}}

\def\alphaox{$\alpha_{\mbox{\tiny OX}}$}
\def\luv{$l_{\mbox{\scriptsize 2500 \AA}}$}
\def\lx{$l_{\mbox{\scriptsize 2 keV}}$}
\def\us{\char`\_}

\def\phn{\phantom{0}}     
\def\phs{\phantom{$-$}}   
\def\simgt{\lower 2pt \hbox{$\, \buildrel {\scriptstyle >}\over {\scriptstyle \sim}\,$}}
\def\simlt{\lower 2pt \hbox{$\, \buildrel {\scriptstyle <}\over {\scriptstyle \sim}\,$}}

\def\chandra{{\slshape Chandra\/}}

\def\heao1{{\slshape HEAO1\/}}
\def\hst{{\slshape HST\/}}

\def\rosat{{\slshape ROSAT\/}}

\def\xmm{{\slshape XMM-Newton\/}}
\def\xray{\mbox{X-ray}}
\shorttitle{The \xray --to--Optical Properties of Optically-Selected Active Galaxies}
\shortauthors{Steffen et al.}
\begin{document}


\title{The X-ray-to-Optical Properties of Optically-Selected Active Galaxies Over Wide Luminosity and Redshift Ranges}

\author{A.~T.~Steffen\altaffilmark{1},
        I.~Strateva\altaffilmark{1},
        W.~N.~Brandt\altaffilmark{1},
        D.~M.~Alexander\altaffilmark{2},
        A.~M.~Koekemoer\altaffilmark{3},
        B.~D.~Lehmer\altaffilmark{1},
        D.~P.~Schneider\altaffilmark{1},
	C.~Vignali\altaffilmark{4,5}}
\altaffiltext{1}{Department of Astronomy and Astrophysics, 525 Davey Laboratory, Pennsylvania State University, University Park, PA 16802.}
\altaffiltext{2}{Institute of Astronomy, Madingley Road, Cambridge CB3 0HA, UK.}
\altaffiltext{3}{Space Telescope Science Institute, 3700 San Martin Drive, Baltimore, MD 21218.}
\altaffiltext{4}{Dipartimento di Astronomia, Universit\`{a} degli Studi di Bologna, Via Ranzani 1, 40127 Bologna, Italy.}
\altaffiltext{5}{INAF-Osservatorio Astronomico di Bologna, Via Ranzani 1, 40127 Bologna, Italy.}

\begin{abstract}
  We present partial-correlation analyses that examine the strengths
  of the relationships between \luv , \lx , \alphaox , and redshift
  for optically-selected AGNs.  We extend the work of
  \citet{strateva05a}, that analyzed optically-selected AGNs from the
  Sloan Digital Sky Survey (SDSS), by including 52
  moderate-luminosity, optically-selected AGNs from the
  \mbox{COMBO-17} survey with corresponding deep ($\approx 250$~ks to
  1~Ms) \xray\ observations from the Extended \chandra\ Deep
  Field-South.  The \mbox{COMBO-17} survey extends $\sim 3$ magnitudes
  deeper than the SDSS and probes the moderate-luminosity AGNs that
  numerically dominate the AGN population in the Universe.  We also
  include recently published observations of 19 high-redshift,
  optically-selected AGNs, and 46 luminous, low-redshift AGNs from the
  Bright Quasar Survey.  The full sample used in our analysis consists
  of 333 AGNs, extending out to $z \sim 6$, with 293 ($88\%$) having
  \xray\ detections. The sample spans five decades in UV luminosity
  and four decades in \xray\ luminosity.  We confirm that \alphaox\ is
  strongly anti-correlated with \luv\ ($13.6\sigma$), the highest
  significance found for this relation to date, and find evidence
  suggesting that the slope of this relation may be dependent on \luv
  .  We find that no significant correlation exists between \alphaox\
  and redshift ($1.3\sigma$), and constrain the maximum evolution of
  AGN UV-to-\xray\ flux ratios to be less than $30\%$ ($1\sigma$) out
  to $z=5$.  Using our sample's high X-ray detection fraction, we also
  find a significant anti-correlation ($3.0\sigma$) between \alphaox\
  and \lx .  We make comparisons to earlier studies on this topic and
  discuss implications for \xray\ vs. optical luminosity functions.
\end{abstract}



\section{Introduction}

The relationship between \xray\ and UV luminosity for AGNs, defined by
\citet{tananbaum79} as $\alpha_{\mbox{\tiny OX}} = {\rm
  log}[l(\nu_{\mbox{\tiny X-ray}})/l(\nu_{\mbox{\tiny UV}})] / {\rm
  log}(\nu_{\mbox{\tiny X-ray}}/\nu_{\mbox{\tiny UV}})$, where
$l(\nu)$ is the monochromatic luminosity (erg~s$^{-1}$~Hz$^{-1}$) at
frequency $\nu$ in the rest-frame, has been the subject of many
studies over the last two decades
\citep[e.g.,][]{avni82,kriss85,avni86,anderson87,wilkes94,vignali03,strateva05a}.
Most of these investigations examined the relationship between the
2500~\AA\ and 2~keV monochromatic luminosities, reducing the \alphaox\
relationship to $\alpha_{\mbox{\tiny OX}} = 0.3838 \:\mbox{log}($\lx
$/$ \luv$)$. These studies have usually found that \alphaox\ is
anti-correlated with UV luminosity, with at most a weak
anti-correlation with redshift \citep[but see][]{yuan98,bechtold03}.
Understanding the relationship between the intrinsic UV and \xray\
emission from AGNs and its evolution through cosmic time is important
for testing energy generation models in AGNs, deriving bolometric
corrections for AGNs which are used in estimating their luminosities
and accretion rates, identifying \xray\ weak AGNs, and estimating
additional \xray\ emission linked with jets in radio-loud AGNs.  In
addition, understanding how \alphaox\ evolves with redshift aids in
resolving discrepancies between the luminosity functions derived from
optical \citep[e.g.,][]{boyle00,wolf03b,richards05} and \xray\
\citep[e.g.,][]{miyaji00,steffen03,ueda03,barger05a,hasinger05,lafranca05}
AGN samples, which map the evolution of the accretion onto
supermassive black-holes (SMBHs) in the Universe.

To provide tight constraints upon \alphaox$(l,z)$, any degeneracies
between luminosity and redshift must be broken.  This is difficult in
flux-limited samples, as $l$ and $z$ are typically strongly
correlated.  To obtain good coverage of the $l-z$ plane while
maintaining a high \xray\ detection fraction, it is necessary to
combine both wide-field and deeper, narrow-field surveys.  To achieve
this goal we expand here upon the \alphaox\ work of \citet[][hereafter
S05]{strateva05a}, who used the wide-field Sloan Digital Sky Survey
\citep[SDSS;][]{york00} quasar sample along with low-redshift
($z<0.2$) Seyfert 1s and high-redshift ($z>4$) optically selected
AGNs.  We add to this sample moderate-luminosity, optically-selected
AGNs from the deep, narrow-field ($0.26$ deg$^{2}$) \mbox{COMBO-17}
survey \citep{wolf04} centered on the Extended \chandra\ Deep
Field-South \citep[\mbox{E-CDF-S;}][]{lehmer05b}, along with a
well-studied sample of luminous Bright Quasar Survey (BQS) quasars
\citep{schmidt83} at $z<0.5$ and additional optically-selected AGNs at
$z>4$ \citep{kelly05,shemmer05a,vignali05a}.  The AGNs optically
selected in the \mbox{COMBO-17} survey extend $\sim3$ magnitudes
fainter than those used in previous \alphaox\ studies (e.g., $B=19.2$;
\citealt{avni86} and $B=19.5$; \citealt{anderson87}) and have a
substantially higher sky density than the SDSS quasar sample
\citep[$\sim 670~{\rm deg}^{-2}$ versus $\sim10~{\rm
  deg}^{-2}$;][]{wolf04,schneider05}.  These \mbox{COMBO-17} AGNs are
representative of most of the optically-selectable AGNs in the
Universe.

While deep \xray\ surveys can provide even higher AGN sky densities
\citep[up to $\simeq 7000$ deg$^{-2}$; e.g.,][]{alexander03}, \xray\
selected samples are dominated by obscured sources
\citep[e.g.,][]{bauer04}.  In the UV, this obscuration leads to
increased contamination from the host galaxy and uncertain absorption
corrections \citep*[e.g.,][]{moran02}.  The \xray\ absorption
corrections are also often uncertain, mainly due to the poor
signal-to-noise ratio of \xray\ spectra with low photon counts.  It is
therefore difficult to measure the {\slshape intrinsic} emission from
these obscured AGNs, and thus they do not significantly elucidate the
intrinsic energy generation mechanisms within AGNs.

The exclusion of obscured sources from our analyses does not
necessarily mean our conclusions below are valid for unobscured
sources only.  Our results are still applicable to obscured AGNs if
the obscuration is a line-of-sight orientation effect and does not
affect the intrinsic energy generation mechanisms within AGNs
\citep[i.e., if the ``Unified'' AGN model is valid; see][for a
review]{antonucci93}.  This will apply even given the recent evidence
for luminosity-dependent obscuration found in \xray\ selected AGN
samples \citep[e.g.,][]{steffen03,ueda03,barger05a,lafranca05}.

We briefly introduce the optical and \xray\ samples used in this paper
in \S2.  In \S3 we discuss the analysis of the correlations between
$l_{\mbox{\scriptsize 2500 \AA}}$, $l_{\mbox{\scriptsize 2~keV}}$,
\alphaox , and redshift.  We discuss our findings and present our
conclusions in \S4.  We use J2000 coordinates and the cosmological
parameters $H_{0} = 70$ km s$^{-1}$ Mpc$^{-1}$, $\Omega_{M} = 0.3$,
and $\Omega_{\Lambda} = 0.7$.

\section{Sample}
To search for correlations among $l_{\mbox{\scriptsize 2500 \AA}}$,
$l_{\mbox{\scriptsize 2 keV}}$, \alphaox , and redshift within
optically-selected AGN samples, it is best to cover as much of the
$l-z$ plane as possible to minimize degeneracies.  It is also
important to eliminate AGNs with intrinsic absorption that can affect
UV and \xray\ photons differently, and radio-loud AGNs that can have
additional UV and \xray\ flux associated with the radio jet
\citep[e.g.,][]{worrall87,wilkes87,worrall05}.  To this end, we
combine the data from S05 with the optically-selected,
moderate-luminosity AGNs from the \mbox{COMBO-17} survey, a subsample
of sources from the BQS quasar catalog which includes
higher-luminosity, optically-selected AGNs, and additional
optically-selected, $z>4$ AGNs.  We describe our AGN sample below
starting with a brief description of the S05 catalog.

\subsection{Strateva et al. (2005) Sample}
The main S05 sample is composed of broad-line AGNs optically selected
via multi-band photometry in the SDSS Data Release~2 \citep[SDSS
DR2;][]{abazajian04}, excluding broad absorption-line quasars
(BALQSOs) and radio-loud sources.  The S05 sample is an unbiased
subsample of the full SDSS DR2 AGN catalog, selecting sources with
medium-deep ($>11$ ks) \rosat\ PSPC observations.  Of the 155 SDSS DR2
AGNs in the S05 sample, 126 ($81\%$) have \xray\ detections.  \xray\
upper limits were calculated for the remaining 29 sources.  The 155
AGNs span redshifts $0.2<z<3.5$.

To extend their $l-z$ plane coverage, S05 include the low-redshift
Seyfert 1 sample of \citet{walter93}, excluding the radio-loud sources
($L_{5 {\rm GHz}} > 10^{25}$ W Hz$^{-1}$).  Twelve of these AGNs are
also included in the well-defined BQS (see \S~2.2.2), and thus we have
removed them from the S05 sample.  For their high-redshift sample, S05
include optically-selected, $z>4$ quasars from the SDSS, the Palomar
Digital Sky Survey \citep{djorgovski98}, and the Automatic Plate
Measuring facility survey \citep{irwin91}.  These high-redshift AGNs
were specifically targeted by \chandra\ and \xmm, and S05 again
eliminated radio-loud sources and BALQSOs from their sample.  It is
important to note that all of these optically-selected, high-redshift
AGNs would have met the SDSS AGN color-selection criteria, so the
introduction of the sources from other surveys does not bias the S05
sample. This high-redshift sample includes 36 AGNs, increasing the S05
total sample size to 216 optically-selected AGNs (after removing the
twelve Seyfert 1s also observed in the BQS), 183 ($85\%$) of which
have detected \xray\ counterparts and 33 ($15\%$) of which have \xray\
upper limits.

\subsection{New Objects in This Study}

\subsubsection{\mbox{COMBO-17}}

%
%
\begin{figure}
\epsscale{1.2}
\plotone{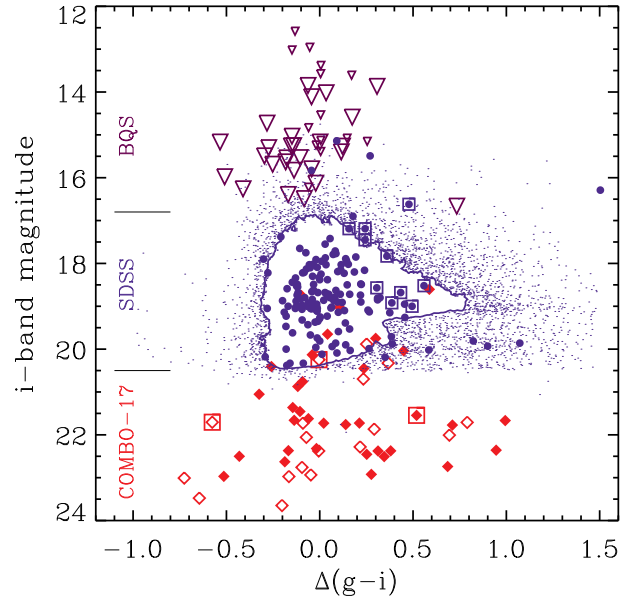} 
\caption{\label{samples}
 Apparent $i$-band PSF magnitude vs. relative $g-i$ color,
 $\Delta(g-i)$, for our selected \mbox{COMBO-17} sample ({\slshape
 red diamonds}), the SDSS sample ({\slshape blue circles}), and the BQS sample
 ({\slshape violet downward-pointing triangles}).  For comparison the full
 SDSS DR2 sample is shown as a blue contour enclosing $90\%$ of the data
 and small blue dots representing the outliers.  Any sources with
 significant galaxy contributions that could affect their observed
 colors (identified spectroscopically for SDSS AGNs and labeled `QSO
 (GAL?)' in the \mbox{COMBO-17} survey) are enclosed by large squares.
 \mbox{COMBO-17} sources with $z>2.1$ are denoted by open diamonds.
 The BQS sample sources observed in the SDSS are denoted with
 large symbols, and sources with SDSS magnitudes derived from other
 bands are denoted with small symbols.  In the computation of
 $\Delta(g-i)$, only AGNs with point-source morphology were used to
 determine the median $g-i$ color as a function of redshift to prevent
 artificial reddening due to host-galaxy contamination.  The
 characteristic $i$-band magnitude ranges for the BQS, SDSS, and
 \mbox{COMBO-17} samples are denoted along the left-hand side of the
 figure.}
\end{figure}

To extend the AGN sample to fainter luminosities we include the
optically-selected AGNs from the portion of the \mbox{COMBO-17} survey
that covers the \mbox{E-CDF-S} \citep{wolf04}.  To determine if the
\mbox{COMBO-17} AGNs are indeed consistent with being a fainter
extension of the SDSS AGNs, we examine the color-magnitude
distributions of the two populations.  Figure~\ref{samples} shows the
apparent $i$-band PSF magnitudes versus the relative $g-i$ colors for
the \mbox{COMBO-17} AGNs ({\slshape diamonds}), the SDSS AGNs
({\slshape circles}), and the BQS sample covered by SDSS
({\slshape large triangles}) and with SDSS colors derived from APM
magnitudes ({\slshape small triangles}).  The SDSS $g$ and $i$
magnitudes were calculated for the \mbox{COMBO-17} AGNs using the
observed $BVRI$ magnitudes and the color transformations calculated by
\citet{jester05a}. These AGN color transformations are for sources
with redshifts $z<2.1$.  In Figure~\ref{samples} we denote
\mbox{COMBO-17} sources beyond this redshift with open diamonds.  The
relative $g-i$ colors, denoted as $\Delta(g-i)$, were constructed by
subtracting the median $g-i$ color of the DR2 AGNs as a function of
redshift from the observed $g-i$ color.  Relative (i.e.,
redshift-corrected) colors can be used to detect significant
differences in the optical/UV continuua of AGNs \citep{richards03}, so
they can be used to compare the SDSS and \mbox{COMBO-17} AGN
populations.  From Figure~\ref{samples} it appears that the
\mbox{COMBO-17} AGN sample is plausibly the faint extension of the
SDSS AGN population.

The \mbox{COMBO-17} survey used observations in 5 broad-band and 12
medium-band filters spanning the optical regime to derive photometric
redshifts and ``fuzzy'' spectra for over $63\,000$ sources in the
\mbox{E-CDF-S} field.  \citet{wolf04} find 175 sources with
photometric colors that are best matched by the SDSS broad-line QSO
template spectrum of \citet{vanden_berk01}.  Their method
preferentially selects broad-line AGNs, but narrow-line sources with
strong AGN emission lines could also be selected.  

The 17-band photometry and use of multiple galaxy/AGN templates allows
the \mbox{COMBO-17} survey to measure among the most accurate
photometric redshifts to date.  To examine the quality of the
photometric redshifts for the \mbox{COMBO-17} AGNs, we compared the
photometric redshifts with a compilation of publicly available
spectroscopic redshifts of sources within the GOODS-South field
created by Alessandro Rettura.\footnotemark\ For our study, we find
spectroscopic redshifts for 12 \mbox{COMBO-17} AGNs and use
photometric redshifts for the remaining sources.  Overall, we find
excellent agreement between the photometric and spectroscopic
redshifts for AGNs at the magnitudes we consider below.  The
spectroscopic redshifts for 11 sources differ by less than $5\%$ from
their measured photometric redshifts.  Only one of the twelve COMBO-17
sources with measured spectroscopic redshifts has a photometric
redshift that significantly disagrees with the spectroscopic value
(COMBO-17 ID: 30792; $z_{\rm phot} = 1.929$; $z_{\rm spec} = 0.743$).

To obtain monochromatic 2 keV \xray\ luminosities (or upper limits)
for these \mbox{COMBO-17} AGNs, we match the sources to the
\mbox{E-CDF-S} and the \chandra\ Deep Field-South
\citep[CDF-S;][]{alexander03} X-ray catalogs.  The \mbox{E-CDF-S} is
composed of four 250~ks \chandra\ observations that cover the entire
\mbox{COMBO-17} field \citep{lehmer05b}.  The 1~Ms CDF-S probes
fainter fluxes than the \mbox{E-CDF-S}, but it does not cover the
entire \mbox{COMBO-17} field.  We use a 2\arcsec\ matching radius to
identify the X-ray counterparts of the COMBO-17 AGNs, and we estimate
the false-match probability to be $<0.5\%$.  All of the
\mbox{COMBO-17} sources with \xray\ counterparts were detected in the
E-CDF-S, and none had more than one X-ray counterpart.  We use the
CDF-S to derive \xray\ upper limits only when \mbox{COMBO-17} AGNs are
not detected in the overlapping \mbox{E-CDF-S} fields.

%
%
\begin{figure}
\includegraphics[angle=90, scale=0.35]{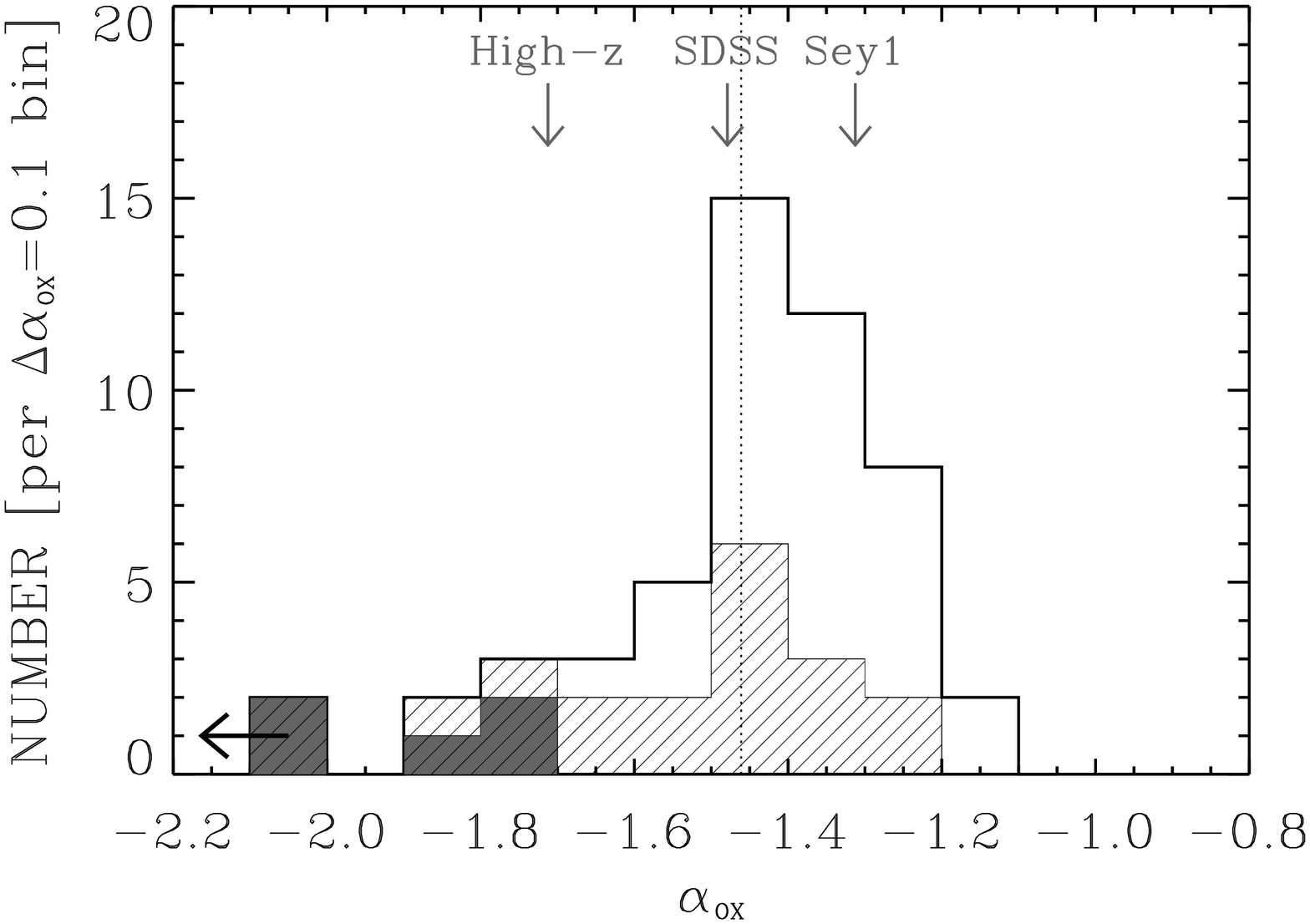}
\caption{\label{alphaox_hist} Histogram of the \alphaox\ values for
  the 52 \mbox{COMBO-17} AGNs included in our sample.  The mean of the
  distribution is shown with a dotted line.  \mbox{COMBO-17} AGNs with
  $\Gamma<1.6$ are marked ({\slshape hatched histogram}).  Sources
  with \alphaox\ upper-limits (i.e., without \xray\ detections) are
  also marked ({\slshape shaded histogram}). The left-pointing arrow
  indicates the direction of the limits.  For comparison, we show the
  mean \alphaox\ values for the three AGN samples in S05, which are described in
  Table~\ref{sample_numbers}.}
\end{figure}

\tabletypesize{\scriptsize}
\setlength{\tabcolsep}{3.75pt}
\begin{deluxetable*}{lcccrccccccc}
  \setlength{\tabcolsep}{1pt}
  \tabletypesize{\scriptsize}
  \tablewidth{0pt}
  \tablecolumns{12}
  \tablecaption{\label{c17_summary}\mbox{COMBO-17}/E-CDF-S AGN DATA.}
  \tablehead{
    \colhead{\mbox{COMBO-17}} &
    \colhead{\mbox{E-CDF-S}} &
    \colhead{$\alpha_{J2000}$} &
    \colhead{$\delta_{J2000}$} &
    \colhead{$z$} &
    \colhead{log$(f_{\mbox{2500 \AA}})$} &
    \colhead{log$(l_{\mbox{2500 \AA}})$} &
    \colhead{log$(f_{\mbox{2 keV}})$} &
    \colhead{log$(l_{\mbox{2 keV}})$} &
    \colhead{\alphaox} &
    \colhead{$\Gamma$} &
    \colhead{$R$} \\
    \colhead{ID} &
    \colhead{ID} &
    \colhead{} &
    \colhead{} &
    \colhead{} &
    \colhead{{\tiny [erg cm$^{-2}$ s$^{-1}$ Hz$^{-1}$]}} &
    \colhead{{\tiny [erg s$^{-1}$ Hz$^{-1}$]}} &
    \colhead{{\tiny [erg cm$^{-2}$ s$^{-1}$ Hz$^{-1}$]}} &
    \colhead{{\tiny [erg s$^{-1}$ Hz$^{-1}$]}} &
    \colhead{} &
    \colhead{} &
    \colhead{} \\
    \colhead{(1)} &
    \colhead{(2)} &
    \colhead{(3)} &
    \colhead{(4)} &
    \colhead{(5)} &
    \colhead{(6)} &
    \colhead{(7)} &
    \colhead{(8)} &
    \colhead{(9)} &
    \colhead{(10)} &
    \colhead{(11)} &
    \colhead{(12)}
  }
  \startdata      
\phn1257 & 399 & $03 \; 32 \; 32.28$ & $-28 \; 03 \; 28.2$ & $1.231$\phantom{\tablenotemark{c}} & \phm{$<$}$\;-28.19$ & \phm{$<$}$\;29.40$ & \phm{$<$}$\;-31.51$ & \phm{$<$}$\;26.08$ & \phm{$<$}$\;-1.27$ & $1.49$ & $<$\phn3.20  \\
\phn1731 & 678 & $03 \; 33 \; 22.85$ & $-28 \; 03 \; 12.9$ & $1.499$\phantom{\tablenotemark{c}} & \phm{$<$}$\;-28.47$ & \phm{$<$}$\;29.29$ & \phm{$<$}$\;-32.20$ & \phm{$<$}$\;25.56$ & \phm{$<$}$\;-1.43$ & $2.04$ & $<$\phn5.79  \\
\phn2006 & 397 & $03 \; 32 \; 32.00$ & $-28 \; 03 \; 09.9$ & $1.966$\phantom{\tablenotemark{c}} & \phm{$<$}$\;-27.37$ & \phm{$<$}$\;30.60$ & \phm{$<$}$\;-31.12$ & \phm{$<$}$\;26.85$ & \phm{$<$}$\;-1.44$ & $1.56$ & $<$\phn0.43  \\
\phn4050 & 357 & $03 \; 32 \; 20.31$ & $-28 \; 02 \; 14.7$ & $1.635$\phantom{\tablenotemark{c}} & \phm{$<$}$\;-27.80$ & \phm{$<$}$\;30.03$ & \phm{$<$}$\;-31.53$ & \phm{$<$}$\;26.29$ & \phm{$<$}$\;-1.43$ & $1.89$ & $<$\phn1.17  \\
\phn4809 & \phn99 & $03 \; 31 \; 36.25$ & $-28 \; 01 \; 49.6$ & $1.988$\phantom{\tablenotemark{c}} & \phm{$<$}$\;-28.31$ & \phm{$<$}$\;29.67$ & \phm{$<$}$\;-31.72$ & \phm{$<$}$\;26.26$ & \phm{$<$}$\;-1.31$ & $1.79$ & $<$\phn3.51  \\
\phn4995 & 234 & $03 \; 31 \; 56.88$ & $-28 \; 01 \; 49.1$ & $1.412$\phantom{\tablenotemark{c}} & \phm{$<$}$\;-28.50$ & \phm{$<$}$\;29.21$ & \phm{$<$}$\;-32.39$ & \phm{$<$}$\;25.31$ & \phm{$<$}$\;-1.49$ & $1.87$ & $<$\phn5.65  \\
\phn5498\tablenotemark{a} & 630 & $03 \; 33 \; 16.07$ & $-28 \; 01 \; 31.3$ & $2.075$\phantom{\tablenotemark{c}} & \phm{$<$}$\;-28.45$ & \phm{$<$}$\;29.56$ & \phm{$<$}$\;-31.97$ & \phm{$<$}$\;26.04$ & \phm{$<$}$\;-1.35$ & $1.62$ & $<$\phn7.22 \\
\phn6735 & 566 & $03 \; 33 \; 06.26$ & $-28 \; 00 \; 55.5$ & $2.444$\phantom{\tablenotemark{c}} & \phm{$<$}$\;-28.32$ & \phm{$<$}$\;29.82$ & \phm{$<$}$\;-31.99$ & \phm{$<$}$\;26.14$ & \phm{$<$}$\;-1.41$ & $1.89$ & $<$\phn3.91  \\
\phn6817 & \phn53 & $03 \; 31 \; 27.79$ & $-28 \; 00 \; 51.0$ & $1.988$\phantom{\tablenotemark{c}} & \phm{$<$}$\;-28.04$ & \phm{$<$}$\;29.94$ & \phm{$<$}$\;-31.96$ & \phm{$<$}$\;26.02$ & \phm{$<$}$\;-1.51$ & $1.49$ & $<$\phn2.33  \\
\phn7570\tablenotemark{b} & \nodata & $03 \; 33 \; 08.03$ & $-28 \; 00 \; 30.8$ & $3.583$\phantom{\tablenotemark{c}} & \phm{$<$}$\;-27.44$ & \phm{$<$}$\;30.97$ & $<-32.81$ & $<25.60$ & $<-2.06$ & \nodata & $<$\phn4.18 \\
\phn7671 & 198 & $03 \; 31 \; 51.80$ & $-28 \; 00 \; 25.7$ & $2.436$\phantom{\tablenotemark{c}} & \phm{$<$}$\;-28.44$ & \phm{$<$}$\;29.69$ & \phm{$<$}$\;-31.97$ & \phm{$<$}$\;26.17$ & \phm{$<$}$\;-1.35$ & $1.94$ & $<$\phn4.47  \\
11818 & \nodata & $03 \; 32 \; 55.98$ & $-27 \; 58 \; 45.3$ & $4.021$\phantom{\tablenotemark{c}} & \phm{$<$}$\;-28.26$ & \phm{$<$}$\;30.23$ & $<-33.47$ & $<25.53$ & $<-1.80$ & \nodata & $<$36.14  \\
11922 & 587 & $03 \; 33 \; 09.11$ & $-27 \; 58 \; 26.6$ & $2.539$\phantom{\tablenotemark{c}} & \phm{$<$}$\;-28.33$ & \phm{$<$}$\;29.84$ & \phm{$<$}$\;-32.02$ & \phm{$<$}$\;26.45$ & \phm{$<$}$\;-1.30$ & $1.42$ & $<$\phn6.67  \\
11941 & 699 & $03 \; 33 \; 26.24$ & $-27 \; 58 \; 29.7$ & $2.172$\phantom{\tablenotemark{c}} & \phm{$<$}$\;-27.74$ & \phm{$<$}$\;30.30$ & \phm{$<$}$\;-31.39$ & \phm{$<$}$\;26.46$ & \phm{$<$}$\;-1.48$ & $2.12$ & $<$\phn1.07  \\
12325 & 541 & $03 \; 33 \; 01.70$ & $-27 \; 58 \; 18.8$ & $1.843$\phantom{\tablenotemark{c}} & \phm{$<$}$\;-27.64$ & \phm{$<$}$\;30.28$ & \phm{$<$}$\;-31.17$ & \phm{$<$}$\;26.66$ & \phm{$<$}$\;-1.39$ & $1.51$ & $<$\phn0.66  \\
13332 & 600 & $03 \; 33 \; 10.63$ & $-27 \; 57 \; 48.5$ & $1.602$\phantom{\tablenotemark{c}} & \phm{$<$}$\;-27.99$ & \phm{$<$}$\;29.82$ & \phm{$<$}$\;-31.53$ & \phm{$<$}$\;26.19$ & \phm{$<$}$\;-1.39$ & $1.97$ & $<$\phn2.11  \\
15396 & 343 & $03 \; 32 \; 16.13$ & $-27 \; 56 \; 44.1$ & $2.682$\phantom{\tablenotemark{c}} & \phm{$<$}$\;-28.39$ & \phm{$<$}$\;29.81$ & \phm{$<$}$\;-32.68$ & \phm{$<$}$\;25.64$ & \phm{$<$}$\;-1.60$ & $1.33$ & $<$\phn6.72  \\
15731 & 711 & $03 \; 33 \; 28.93$ & $-27 \; 56 \; 41.1$ & $0.835$\phantom{\tablenotemark{c}} & \phm{$<$}$\;-27.65$ & \phm{$<$}$\;29.61$ & \phm{$<$}$\;-30.82$ & \phm{$<$}$\;26.29$ & \phm{$<$}$\;-1.28$ & $1.82$ & $<$\phn0.78  \\
16621 & 595 & $03 \; 33 \; 09.70$ & $-27 \; 56 \; 13.9$ & $2.540$\phantom{\tablenotemark{c}} & \phm{$<$}$\;-27.47$ & \phm{$<$}$\;30.69$ & \phm{$<$}$\;-32.05$ & \phm{$<$}$\;26.44$ & \phm{$<$}$\;-1.63$ & $1.80$ & $<$\phn0.60  \\
17229 & \nodata & $03 \; 32 \; 22.57$ & $-27 \; 55 \; 54.4$ & $2.870$\phantom{\tablenotemark{c}} & \phm{$<$}$\;-28.54$ & \phm{$<$}$\;29.71$ & $<-33.20$ & $<25.07$ & $<-1.78$ & \nodata & $<$14.68  \\
18256 & 677 & $03 \; 33 \; 22.79$ & $-27 \; 55 \; 23.7$ & $1.647$\phantom{\tablenotemark{c}} & \phm{$<$}$\;-27.78$ & \phm{$<$}$\;30.05$ & \phm{$<$}$\;-31.93$ & \phm{$<$}$\;26.06$ & \phm{$<$}$\;-1.53$ & $2.01$ & $<$\phn1.14  \\
18324\tablenotemark{a} & 531 & $03 \; 33 \; 00.77$ & $-27 \; 55 \; 20.6$ & $1.990$\phantom{\tablenotemark{c}} & \phm{$<$}$\;-28.43$ & \phm{$<$}$\;29.55$ & \phm{$<$}$\;-31.43$ & \phm{$<$}$\;26.47$ & \phm{$<$}$\;-1.18$ & $2.23$ & $<$\phn6.82  \\
19965 & 154 & $03 \; 31 \; 45.20$ & $-27 \; 54 \; 35.6$ & $0.634$\phantom{\tablenotemark{c}} & \phm{$<$}$\;-27.51$ & \phm{$<$}$\;29.51$ & \phm{$<$}$\;-31.22$ & \phm{$<$}$\;25.47$ & \phm{$<$}$\;-1.55$ & $1.92$ & $<$\phn0.55  \\
25042 & 434 & $03 \; 32 \; 41.85$ & $-27 \; 52 \; 02.4$ & $3.610$\tablenotemark{c} & \phm{$<$}$\;-28.08$ & \phm{$<$}$\;30.34$ & \phm{$<$}$\;-32.02$ & \phm{$<$}$\;26.48$ & \phm{$<$}$\;-1.48$ & $1.43$ & $<$12.09  \\
25884 & \phn95 & $03 \; 31 \; 35.77$ & $-27 \; 51 \; 34.7$ & $1.630$\phantom{\tablenotemark{c}} & \phm{$<$}$\;-28.14$ & \phm{$<$}$\;29.68$ & \phm{$<$}$\;-31.43$ & \phm{$<$}$\;26.39$ & \phm{$<$}$\;-1.26$ & $1.67$ & $<$\phn3.61  \\
30792 & 440 & $03 \; 32 \; 43.24$ & $-27 \; 49 \; 14.1$ & $0.743$\tablenotemark{c} & \phm{$<$}$\;-28.60$ & \phm{$<$}$\;28.56$ & \phm{$<$}$\;-31.79$ & \phm{$<$}$\;25.14$ & \phm{$<$}$\;-1.31$ & $1.66$ & $<$\phn5.72  \\
33069 & 308 & $03 \; 32 \; 09.45$ & $-27 \; 48 \; 06.7$ & $2.810$\tablenotemark{c} & \phm{$<$}$\;-27.53$ & \phm{$<$}$\;30.71$ & \phm{$<$}$\;-32.23$ & \phm{$<$}$\;26.04$ & \phm{$<$}$\;-1.79$ & $0.82$ & $<$\phn1.41  \\
33630 & 119 & $03 \; 31 \; 40.12$ & $-27 \; 47 \; 46.4$ & $2.719$\phantom{\tablenotemark{c}} & \phm{$<$}$\;-28.32$ & \phm{$<$}$\;29.90$ & \phm{$<$}$\;-31.99$ & \phm{$<$}$\;26.23$ & \phm{$<$}$\;-1.41$ & $1.59$ & $<$\phn3.72  \\
33644 & 525 & $03 \; 32 \; 59.85$ & $-27 \; 47 \; 48.2$ & $2.538$\phantom{\tablenotemark{c}} & \phm{$<$}$\;-28.15$ & \phm{$<$}$\;30.02$ & \phm{$<$}$\;-31.74$ & \phm{$<$}$\;26.76$ & \phm{$<$}$\;-1.25$ & $1.71$ & $<$\phn2.35  \\
34357 & 304 & $03 \; 32 \; 08.67$ & $-27 \; 47 \; 34.2$ & $0.543$\tablenotemark{c} & \phm{$<$}$\;-27.33$ & \phm{$<$}$\;29.55$ & \phm{$<$}$\;-30.55$ & \phm{$<$}$\;25.89$ & \phm{$<$}$\;-1.40$ & $1.73$ & \phs13.89  \\
36120 & 180 & $03 \; 31 \; 49.41$ & $-27 \; 46 \; 34.0$ & $2.306$\phantom{\tablenotemark{c}} & \phm{$<$}$\;-28.35$ & \phm{$<$}$\;29.74$ & \phm{$<$}$\;-32.09$ & \phm{$<$}$\;25.85$ & \phm{$<$}$\;-1.49$ & $1.42$ & $<$\phn5.10  \\
37487 & 422 & $03 \; 32 \; 39.08$ & $-27 \; 46 \; 01.8$ & $1.216$\tablenotemark{c} & \phm{$<$}$\;-27.86$ & \phm{$<$}$\;29.73$ & \phm{$<$}$\;-32.68$ & \phm{$<$}$\;25.01$ & \phm{$<$}$\;-1.81$ & $1.07$ & $<$\phn1.25  \\
38551 & 390 & $03 \; 32 \; 29.98$ & $-27 \; 45 \; 29.8$ & $1.221$\tablenotemark{c} & \phm{$<$}$\;-28.03$ & \phm{$<$}$\;29.56$ & \phm{$<$}$\;-31.53$ & \phm{$<$}$\;26.00$ & \phm{$<$}$\;-1.37$ & $1.66$ & $<$\phn1.88  \\
38905 & 549 & $03 \; 33 \; 03.62$ & $-27 \; 45 \; 18.7$ & $1.264$\phantom{\tablenotemark{c}} & \phm{$<$}$\;-28.71$ & \phm{$<$}$\;28.90$ & \phm{$<$}$\;-32.05$ & \phm{$<$}$\;25.66$ & \phm{$<$}$\;-1.25$ & $1.55$ & $<$11.94  \\
39432 & 393 & $03 \; 32 \; 30.22$ & $-27 \; 45 \; 04.5$ & $0.735$\tablenotemark{c} & \phm{$<$}$\;-28.52$ & \phm{$<$}$\;28.63$ & \phm{$<$}$\;-31.85$ & \phm{$<$}$\;25.33$ & \phm{$<$}$\;-1.27$ & $1.88$ & $<$\phn4.83  \\
42601\tablenotemark{b} & 515 & $03 \; 32 \; 59.07$ & $-27 \; 43 \; 39.5$ & $0.510$\phantom{\tablenotemark{c}} & \phm{$<$}$\;-27.98$ & \phm{$<$}$\;28.85$ & \phm{$<$}$\;-31.52$ & \phm{$<$}$\;24.90$ & \phm{$<$}$\;-1.51$ & $1.67$ & $<$\phn3.60  \\
43151 & 249 & $03 \; 32 \; 00.36$ & $-27 \; 43 \; 19.5$ & $1.037$\tablenotemark{c} & \phm{$<$}$\;-28.42$ & \phm{$<$}$\;29.03$ & \phm{$<$}$\;-31.70$ & \phm{$<$}$\;25.48$ & \phm{$<$}$\;-1.36$ & $1.54$ & $<$\phn4.70  \\
44126 & 583 & $03 \; 33 \; 08.78$ & $-27 \; 42 \; 54.4$ & $0.729$\phantom{\tablenotemark{c}} & \phm{$<$}$\;-28.91$ & \phm{$<$}$\;28.23$ & \phm{$<$}$\;-31.78$ & \phm{$<$}$\;25.06$ & \phm{$<$}$\;-1.22$ & $1.74$ & $<$11.20  \\
47615 & 191 & $03 \; 31 \; 50.96$ & $-27 \; 41 \; 15.7$ & $0.649$\phantom{\tablenotemark{c}} & \phm{$<$}$\;-28.52$ & \phm{$<$}$\;28.52$ & \phm{$<$}$\;-31.96$ & \phm{$<$}$\;25.11$ & \phm{$<$}$\;-1.31$ & $1.79$ & $<$\phn5.92  \\
48284 & 378 & $03 \; 32 \; 27.01$ & $-27 \; 41 \; 04.9$ & $0.734$\tablenotemark{c} & \phm{$<$}$\;-27.16$ & \phm{$<$}$\;29.99$ & \phm{$<$}$\;-30.61$ & \phm{$<$}$\;26.25$ & \phm{$<$}$\;-1.43$ & $1.79$ & \phs\phn5.59  \\
48870 & 712 & $03 \; 33 \; 28.94$ & $-27 \; 40 \; 43.7$ & $2.146$\phantom{\tablenotemark{c}} & \phm{$<$}$\;-28.46$ & \phm{$<$}$\;29.57$ & \phm{$<$}$\;-31.68$ & \phm{$<$}$\;26.66$ & \phm{$<$}$\;-1.12$ & $1.61$ & $<$\phn5.64  \\
49298 & 374 & $03 \; 32 \; 26.49$ & $-27 \; 40 \; 35.5$ & $1.031$\tablenotemark{c} & \phm{$<$}$\;-27.52$ & \phm{$<$}$\;29.93$ & \phm{$<$}$\;-31.37$ & \phm{$<$}$\;26.03$ & \phm{$<$}$\;-1.49$ & $2.08$ & $<$\phn0.61  \\
50415\tablenotemark{a} & 412 & $03 \; 32 \; 37.45$ & $-27 \; 40 \; 00.1$ & $0.666$\tablenotemark{c} & \phm{$<$}$\;-28.89$ & \phm{$<$}$\;28.17$ & \phm{$<$}$\;-32.21$ & \phm{$<$}$\;24.73$ & \phm{$<$}$\;-1.32$ & $1.83$ & $<$11.25 \\
50997 & 416 & $03 \; 32 \; 38.12$ & $-27 \; 39 \; 44.8$ & $0.837$\tablenotemark{c} & \phm{$<$}$\;-27.71$ & \phm{$<$}$\;29.55$ & \phm{$<$}$\;-31.38$ & \phm{$<$}$\;25.90$ & \phm{$<$}$\;-1.40$ & $2.05$ & $<$\phn0.92  \\
51835 & 118 & $03 \; 31 \; 40.05$ & $-27 \; 39 \; 17.6$ & $2.179$\phantom{\tablenotemark{c}} & \phm{$<$}$\;-28.60$ & \phm{$<$}$\;29.45$ & \phm{$<$}$\;-32.51$ & \phm{$<$}$\;25.63$ & \phm{$<$}$\;-1.47$ & $1.50$ & $<$\phn7.73  \\
52963 & \nodata & $03 \; 32 \; 52.60$ & $-27 \; 38 \; 46.2$ & $0.548$\phantom{\tablenotemark{c}} & \phm{$<$}$\;-28.67$ & \phm{$<$}$\;28.22$ & $<-33.30$ & $<23.59$ & $<-1.78$ & \nodata & $<$\phn7.38  \\
54839 & 116 & $03 \; 31 \; 39.25$ & $-27 \; 37 \; 52.0$ & $1.428$\phantom{\tablenotemark{c}} & \phm{$<$}$\;-28.34$ & \phm{$<$}$\;29.38$ & \phm{$<$}$\;-31.88$ & \phm{$<$}$\;26.04$ & \phm{$<$}$\;-1.28$ & $1.88$ & $<$\phn4.61  \\
54969\tablenotemark{b} & \nodata & $03 \; 32 \; 49.28$ & $-27 \; 37 \; 56.7$ & $3.361$\phantom{\tablenotemark{c}} & \phm{$<$}$\;-27.39$ & \phm{$<$}$\;30.97$ & $<-33.09$ & $<25.61$ & $<-2.06$ & \nodata & $<$\phn1.23 \\
56074 & 320 & $03 \; 32 \; 11.65$ & $-27 \; 37 \; 25.9$ & $1.574$\phantom{\tablenotemark{c}} & \phm{$<$}$\;-26.96$ & \phm{$<$}$\;30.83$ & \phm{$<$}$\;-31.13$ & \phm{$<$}$\;26.84$ & \phm{$<$}$\;-1.53$ & $1.44$ & \phs19.50  \\
57653 & \phn35 & $03 \; 31 \; 23.53$ & $-27 \; 36 \; 31.6$ & $1.653$\phantom{\tablenotemark{c}} & \phm{$<$}$\;-28.25$ & \phm{$<$}$\;29.58$ & \phm{$<$}$\;-32.70$ & \phm{$<$}$\;25.25$ & \phm{$<$}$\;-1.66$ & $1.40$ & $<$\phn3.74  \\
58478 & 516 & $03 \; 32 \; 59.19$ & $-27 \; 36 \; 11.7$ & $1.348$\phantom{\tablenotemark{c}} & \phm{$<$}$\;-28.01$ & \phm{$<$}$\;29.66$ & \phm{$<$}$\;-32.05$ & \phm{$<$}$\;25.77$ & \phm{$<$}$\;-1.49$ & $2.12$ & $<$\phn2.55  \\
60939 & 245 & $03 \; 31 \; 58.91$ & $-27 \; 35 \; 16.1$ & $2.794$\phantom{\tablenotemark{c}} & \phm{$<$}$\;-28.29$ & \phm{$<$}$\;29.94$ & \phm{$<$}$\;-32.37$ & \phm{$<$}$\;26.21$ & \phm{$<$}$\;-1.43$ & $0.96$ & $<$\phn7.46 
  \enddata
\tablenotetext{a}{Source is blended with nearby sources in COMBO-17 images.}
\tablenotetext{b}{Source is identified as 'QSO (GAL?)' in the COMBO-17 survey.}
\tablenotetext{c}{Spectroscopic redshift}
\end{deluxetable*}

To minimize potential contamination of our correlation analysis below
we impose two unbiased selection criteria on the full \mbox{COMBO-17}
AGN sample.  First, we only consider sources brighter than $R=23$.  At
these magnitudes the probability of a \mbox{COMBO-17} source being
misidentified as an AGN is low \citep[$\sim 1\%$;][]{wolf04}.  In
addition, at $R>23$ the fraction of \mbox{COMBO-17} AGNs with \xray\
detections drops quickly.  This magnitude threshold ensures that the
fraction of AGNs with \xray\ detections remains high, reducing the
impact of \xray\ limits in our analysis.  Second, we only include
sources that have \hst\ ACS data from either the Galaxy Evolution from
Morphology and SEDs survey \citep[GEMS;][]{rix04} or the Great
Observatories Origins Deep Survey \citep[GOODS;][]{giavalisco04}.
These high-resolution data, which cover $\sim84\%$ of the
\mbox{COMBO-17} field, can be used to minimize the galaxy contribution
to the UV luminosity of the \mbox{COMBO-17} sources identified as
being extended objects \citep[flagged as `QSO (GAL?)'  in][]{wolf04}.
These two selection criteria reduce the \mbox{COMBO-17} AGN sample
from 175 to 60 sources, with 49 ($82\%$) having \xray\ detections and
13 ($22\%$) identified as extended objects.

\footnotetext{Available at {\ttfamily http://www.eso.org/\discretionary{}{}{}science/\discretionary{}{}{}goods/\discretionary{}{}{}spectroscopy/\discretionary{}{}{}CDFS\us Mastercat/}}

Using the high-resolution GEMS and GOODS ACS $V$-band images we
identified four \mbox{COMBO-17} AGNs (\mbox{COMBO-17}
IDs: 5498, 18324, 31898, and 50415) that were incorrectly classified
as extended sources but are actually close sources blended together in
the lower-resolution \mbox{COMBO-17} images.  We correct the
\mbox{COMBO-17} classification for these blended sources.  For the
remaining nine extended \mbox{COMBO-17} AGNs, we attempted to minimize
the contribution of the host galaxies to the measured optical/UV
luminosities of the AGNs.  Using the high-resolution ACS images we
compared the magnitudes of the \mbox{COMBO-17} AGNs using 3\arcsec\
and 0.5\arcsec\ apertures.  A mean magnitude difference of 0.352 mag
was calculated for the unresolved \mbox{COMBO-17} AGNs.  For the
extended sources, we calculated the magnitude differences using the
same aperture sizes.  Any magnitude offsets exceeding the mean
difference calculated for the point-like AGNs were considered to be
caused by excess light from the host galaxy.  We subtracted this
host-galaxy contribution from the observed \mbox{COMBO-17} $R$-band
magnitude and found that six of the nine extended sources subsequently
fell below our $R=23$ magnitude limit.  Since these extended AGNs
would not have met our magnitude criterion without the additional flux
from the host galaxy, we exclude them from our sample, reducing it to
54 sources.  None of these six AGNs has an \xray\ detection.

We attempted to remove both radio-loud and obscured AGNs from our
sample of \mbox{COMBO-17} AGNs.  Using 1.4~GHz VLA radio maps covering
the \mbox{COMBO-17} field \citep{kellermann04} we found that only five
of the 54 \mbox{COMBO-17} AGNs have detected radio counterparts.  We
used the 1.4~GHz flux limit of $40 \mu$Jy ($5\sigma$), quoted by
\citet{kellermann04}, as an upper limit for the undetected radio
sources.  We calculated the radio-loudness parameter, $R$, using the
\mbox{COMBO-17} $B$-band monochromatic flux ($\lambda = 4350$~\AA) and
converting the observed 1.4~GHz flux density to rest-frame 5~GHz,
assuming $S(\nu) \propto \nu^{-0.8}$.  We eliminate the two
\mbox{COMBO-17} AGNs with $R>30$, and give the values (or limits) of
$R$ for the remaining sources in Table~\ref{c17_summary}.  Our final
\mbox{COMBO-17} AGN sample consists of 52 sources, with 47 ($90\%$)
having \xray\ counterparts.

\tabletypesize{\footnotesize}
\begin{deluxetable*}{lllcccccc}
  \tabletypesize{\scriptsize}
  \tablewidth{0pt}
  \tablecolumns{9}
  \tablecaption{\label{bqs_summary}\mbox{BQS}/\rosat AGN DATA.}
  \tablehead{
    \colhead{Name} &
    \colhead{$\alpha_{J2000}$} &
    \colhead{$\delta_{J2000}$} &
    \colhead{$z$} &
    \colhead{log$(f_{\mbox{\scriptsize 2500 \AA}})$} &
    \colhead{log$(l_{\mbox{\scriptsize 2500 \AA}})$} &
    \colhead{log$(f_{\mbox{\scriptsize 2 keV}})$} &
    \colhead{log$(l_{\mbox{\scriptsize 2 keV}})$} &
    \colhead{\alphaox} \\
    \colhead{} &
    \colhead{} &
    \colhead{} &
    \colhead{} &
    \colhead{{\fontsize{6}{8}\selectfont [erg cm$^{-2}$ s$^{-1}$ Hz$^{-1}$]}} &
    \colhead{{\fontsize{6}{8}\selectfont [erg s$^{-1}$ Hz$^{-1}$]}} &
    \colhead{{\fontsize{6}{8}\selectfont [erg cm$^{-2}$ s$^{-1}$ Hz$^{-1}$]}} &
    \colhead{{\fontsize{6}{8}\selectfont [erg s$^{-1}$ Hz$^{-1}$]}} &
    \colhead{} \\
    \colhead{(1)} &
    \colhead{(2)} &
    \colhead{(3)} &
    \colhead{(4)} &
    \colhead{(5)} &
    \colhead{(6)} &
    \colhead{(7)} &
    \colhead{(8)} &
    \colhead{(9)} 
  }
  \startdata      
PG 0026+129 & $00 \; 29 \; 13.8$ & $+13 \; 16 \; 05$ & $0.142$ & \phm{$<$}$\;-25.47$ & \phm{$<$}$\;30.20$ & \phm{$<$}$\;-29.21$ & \phm{$<$}$\;26.46$ & \phm{$<$}$\;-1.44$ \\
PG 0050+124 & $00 \; 53 \; 34.9$ & $+12 \; 41 \; 36$ & $0.061$ & \phm{$<$}$\;-25.24$ & \phm{$<$}$\;29.69$ & \phm{$<$}$\;-28.96$ & \phm{$<$}$\;25.96$ & \phm{$<$}$\;-1.43$ \\
PG 0052+251 & $00 \; 54 \; 52.2$ & $+25 \; 25 \; 39$ & $0.155$ & \phm{$<$}$\;-25.51$ & \phm{$<$}$\;30.24$ & \phm{$<$}$\;-29.12$ & \phm{$<$}$\;26.64$ & \phm{$<$}$\;-1.39$ \\
PG 0157+001 & $01 \; 59 \; 50.25$ & $+00 \; 23 \; 40.8$ & $0.164$ & \phm{$<$}$\;-25.73$ & \phm{$<$}$\;30.07$ & \phm{$<$}$\;-29.83$ & \phm{$<$}$\;25.97$ & \phm{$<$}$\;-1.57$ \\
PG 0804+761 & $08 \; 10 \; 58.5$ & $+76 \; 02 \; 43$ & $0.100$ & \phm{$<$}$\;-25.01$ & \phm{$<$}$\;30.35$ & \phm{$<$}$\;-28.91$ & \phm{$<$}$\;26.45$ & \phm{$<$}$\;-1.50$ \\
PG 0838+770 & $08 \; 44 \; 45.3$ & $+76 \; 53 \; 10$ & $0.131$ & \phm{$<$}$\;-25.79$ & \phm{$<$}$\;29.81$ & \phm{$<$}$\;-29.76$ & \phm{$<$}$\;25.84$ & \phm{$<$}$\;-1.52$ \\
PG 0844+349 & $08 \; 47 \; 42.47$ & $+34 \; 45 \; 04.4$ & $0.064$ & \phm{$<$}$\;-25.40$ & \phm{$<$}$\;29.57$ & \phm{$<$}$\;-30.28$ & \phm{$<$}$\;24.69$ & \phm{$<$}$\;-1.87$ \\
PG 0923+201 & $09 \; 25 \; 54.7$ & $+19 \; 54 \; 04$ & $0.190$ & \phm{$<$}$\;-25.64$ & \phm{$<$}$\;30.29$ & \phm{$<$}$\;-29.59$ & \phm{$<$}$\;26.34$ & \phm{$<$}$\;-1.52$ \\
PG 0947+396 & $09 \; 50 \; 48.39$ & $+39 \; 26 \; 50.5$ & $0.206$ & \phm{$<$}$\;-26.05$ & \phm{$<$}$\;29.96$ & \phm{$<$}$\;-29.60$ & \phm{$<$}$\;26.41$ & \phm{$<$}$\;-1.36$ \\
PG 0953+414 & $09 \; 56 \; 52.39$ & $+41 \; 15 \; 22.2$ & $0.239$ & \phm{$<$}$\;-25.50$ & \phm{$<$}$\;30.64$ & \phm{$<$}$\;-29.50$ & \phm{$<$}$\;26.64$ & \phm{$<$}$\;-1.54$ \\
PG 1012+008 & $10 \; 14 \; 54.90$ & $+00 \; 33 \; 37.4$ & $0.185$ & \phm{$<$}$\;-25.83$ & \phm{$<$}$\;30.08$ & \phm{$<$}$\;-30.02$ & \phm{$<$}$\;25.89$ & \phm{$<$}$\;-1.61$ \\
PG 1048+342 & $10 \; 51 \; 43.8$ & $+33 \; 59 \; 26$ & $0.167$ & \phm{$<$}$\;-25.94$ & \phm{$<$}$\;29.88$ & \phm{$<$}$\;-29.85$ & \phm{$<$}$\;25.97$ & \phm{$<$}$\;-1.50$ \\
PG 1049-005 & $10 \; 51 \; 51.50$ & $-00 \; 51 \; 17.7$ & $0.357$ & \phm{$<$}$\;-25.76$ & \phm{$<$}$\;30.74$ & \phm{$<$}$\;-30.18$ & \phm{$<$}$\;26.32$ & \phm{$<$}$\;-1.70$ \\
PG 1114+445 & $11 \; 17 \; 06.40$ & $+44 \; 13 \; 33.3$ & $0.144$ & \phm{$<$}$\;-25.88$ & \phm{$<$}$\;29.81$ & \phm{$<$}$\;-30.14$ & \phm{$<$}$\;25.54$ & \phm{$<$}$\;-1.64$ \\
PG 1115+407 & $11 \; 18 \; 30.29$ & $+40 \; 25 \; 54.0$ & $0.154$ & \phm{$<$}$\;-26.06$ & \phm{$<$}$\;29.69$ & \phm{$<$}$\;-29.86$ & \phm{$<$}$\;25.89$ & \phm{$<$}$\;-1.46$ \\
PG 1116+215 & $11 \; 19 \; 08.8$ & $+21 \; 19 \; 18$ & $0.177$ & \phm{$<$}$\;-25.26$ & \phm{$<$}$\;30.62$ & \phm{$<$}$\;-29.35$ & \phm{$<$}$\;26.52$ & \phm{$<$}$\;-1.57$ \\
PG 1121+422 & $11 \; 24 \; 39.18$ & $+42 \; 01 \; 45.0$ & $0.234$ & \phm{$<$}$\;-26.15$ & \phm{$<$}$\;29.97$ & \phm{$<$}$\;-29.88$ & \phm{$<$}$\;26.24$ & \phm{$<$}$\;-1.43$ \\
PG 1151+117 & $11 \; 53 \; 49.27$ & $+11 \; 28 \; 30.4$ & $0.176$ & \phm{$<$}$\;-25.90$ & \phm{$<$}$\;29.96$ & \phm{$<$}$\;-29.56$ & \phm{$<$}$\;26.30$ & \phm{$<$}$\;-1.40$ \\
PG 1202+281 & $12 \; 04 \; 42.1$ & $+27 \; 54 \; 12$ & $0.165$ & \phm{$<$}$\;-26.07$ & \phm{$<$}$\;29.73$ & \phm{$<$}$\;-29.52$ & \phm{$<$}$\;26.29$ & \phm{$<$}$\;-1.32$ \\
PG 1211+143 & $12 \; 14 \; 17.7$ & $+14 \; 03 \; 13$ & $0.085$ & \phm{$<$}$\;-25.04$ & \phm{$<$}$\;30.18$ & \phm{$<$}$\;-28.85$ & \phm{$<$}$\;26.37$ & \phm{$<$}$\;-1.46$ \\
PG 1216+069 & $12 \; 19 \; 20.93$ & $+06 \; 38 \; 38.5$ & $0.334$ & \phm{$<$}$\;-25.72$ & \phm{$<$}$\;30.72$ & \phm{$<$}$\;-29.58$ & \phm{$<$}$\;26.86$ & \phm{$<$}$\;-1.48$ \\
PG 1229+204 & $12 \; 32 \; 03.6$ & $+20 \; 09 \; 30$ & $0.064$ & \phm{$<$}$\;-25.48$ & \phm{$<$}$\;29.49$ & \phm{$<$}$\;-29.15$ & \phm{$<$}$\;25.82$ & \phm{$<$}$\;-1.41$ \\
PG 1259+593 & $13 \; 01 \; 12.93$ & $+59 \; 02 \; 06.8$ & $0.472$ & \phm{$<$}$\;-25.69$ & \phm{$<$}$\;31.06$ & $<-30.20$ & $<26.55$ & $<-1.73$ \\
PG 1307+085 & $13 \; 09 \; 47.00$ & $+08 \; 19 \; 48.2$ & $0.155$ & \phm{$<$}$\;-25.54$ & \phm{$<$}$\;30.21$ & \phm{$<$}$\;-29.47$ & \phm{$<$}$\;26.29$ & \phm{$<$}$\;-1.51$ \\
PG 1322+659 & $13 \; 23 \; 49.52$ & $+65 \; 41 \; 48.2$ & $0.168$ & \phm{$<$}$\;-25.83$ & \phm{$<$}$\;29.99$ & \phm{$<$}$\;-29.56$ & \phm{$<$}$\;26.26$ & \phm{$<$}$\;-1.43$ \\
PG 1352+183 & $13 \; 54 \; 35.6$ & $+18 \; 05 \; 18$ & $0.158$ & \phm{$<$}$\;-25.78$ & \phm{$<$}$\;29.99$ & \phm{$<$}$\;-29.59$ & \phm{$<$}$\;26.18$ & \phm{$<$}$\;-1.46$ \\
PG 1354+213 & $13 \; 56 \; 32.8$ & $+21 \; 03 \; 51$ & $0.300$ & \phm{$<$}$\;-26.29$ & \phm{$<$}$\;30.06$ & \phm{$<$}$\;-29.98$ & \phm{$<$}$\;26.37$ & \phm{$<$}$\;-1.42$ \\
PG 1402+261 & $14 \; 05 \; 16.2$ & $+25 \; 55 \; 34$ & $0.164$ & \phm{$<$}$\;-25.62$ & \phm{$<$}$\;30.19$ & \phm{$<$}$\;-29.45$ & \phm{$<$}$\;26.35$ & \phm{$<$}$\;-1.47$ \\
PG 1404+226 & $14 \; 06 \; 21.9$ & $+22 \; 23 \; 47$ & $0.098$ & \phm{$<$}$\;-26.10$ & \phm{$<$}$\;29.25$ & \phm{$<$}$\;-29.51$ & \phm{$<$}$\;25.83$ & \phm{$<$}$\;-1.31$ \\
PG 1415+451 & $14 \; 17 \; 00.82$ & $+44 \; 56 \; 06.4$ & $0.114$ & \phm{$<$}$\;-25.95$ & \phm{$<$}$\;29.52$ & \phm{$<$}$\;-29.97$ & \phm{$<$}$\;25.51$ & \phm{$<$}$\;-1.54$ \\
PG 1416-129 & $14 \; 19 \; 03.8$ & $-13 \; 10 \; 45$ & $0.129$ & \phm{$<$}$\;-25.12$ & \phm{$<$}$\;30.47$ & \phm{$<$}$\;-29.17$ & \phm{$<$}$\;26.42$ & \phm{$<$}$\;-1.55$ \\
PG 1426+015 & $14 \; 29 \; 06.57$ & $+01 \; 17 \; 06.1$ & $0.086$ & \phm{$<$}$\;-25.13$ & \phm{$<$}$\;30.10$ & \phm{$<$}$\;-28.91$ & \phm{$<$}$\;26.32$ & \phm{$<$}$\;-1.45$ \\
PG 1427+480 & $14 \; 29 \; 43.07$ & $+47 \; 47 \; 26.2$ & $0.221$ & \phm{$<$}$\;-26.09$ & \phm{$<$}$\;29.98$ & \phm{$<$}$\;-29.96$ & \phm{$<$}$\;26.11$ & \phm{$<$}$\;-1.49$ \\
PG 1435-067 & $14 \; 38 \; 16.2$ & $-06 \; 58 \; 20$ & $0.129$ & \phm{$<$}$\;-25.39$ & \phm{$<$}$\;30.20$ & \phm{$<$}$\;-29.47$ & \phm{$<$}$\;26.12$ & \phm{$<$}$\;-1.56$ \\
PG 1440+356 & $14 \; 42 \; 07.47$ & $+35 \; 26 \; 23.0$ & $0.077$ & \phm{$<$}$\;-25.52$ & \phm{$<$}$\;29.61$ & \phm{$<$}$\;-29.03$ & \phm{$<$}$\;26.10$ & \phm{$<$}$\;-1.35$ \\
PG 1444+407 & $14 \; 46 \; 45.94$ & $+40 \; 35 \; 05.8$ & $0.267$ & \phm{$<$}$\;-25.85$ & \phm{$<$}$\;30.39$ & \phm{$<$}$\;-29.98$ & \phm{$<$}$\;26.26$ & \phm{$<$}$\;-1.59$ \\
PG 1519+226 & $15 \; 21 \; 14.3$ & $+22 \; 27 \; 44$ & $0.137$ & \phm{$<$}$\;-25.86$ & \phm{$<$}$\;29.78$ & \phm{$<$}$\;-29.95$ & \phm{$<$}$\;25.69$ & \phm{$<$}$\;-1.57$ \\
PG 1543+489 & $15 \; 45 \; 30.24$ & $+48 \; 46 \; 09.1$ & $0.400$ & \phm{$<$}$\;-26.00$ & \phm{$<$}$\;30.61$ & \phm{$<$}$\;-30.47$ & \phm{$<$}$\;26.14$ & \phm{$<$}$\;-1.72$ \\
PG 1552+085 & $15 \; 54 \; 44.58$ & $+08 \; 22 \; 21.5$ & $0.119$ & \phm{$<$}$\;-25.81$ & \phm{$<$}$\;29.71$ & \phm{$<$}$\;-30.13$ & \phm{$<$}$\;25.38$ & \phm{$<$}$\;-1.66$ \\
PG 1612+261 & $16 \; 14 \; 13.20$ & $+26 \; 04 \; 16.2$ & $0.131$ & \phm{$<$}$\;-25.77$ & \phm{$<$}$\;29.83$ & \phm{$<$}$\;-29.15$ & \phm{$<$}$\;26.45$ & \phm{$<$}$\;-1.30$ \\
PG 1613+658 & $16 \; 13 \; 57.2$ & $+65 \; 43 \; 10$ & $0.129$ & \phm{$<$}$\;-25.78$ & \phm{$<$}$\;29.81$ & \phm{$<$}$\;-29.09$ & \phm{$<$}$\;26.50$ & \phm{$<$}$\;-1.27$ \\
PG 1617+175 & $16 \; 20 \; 11.2$ & $+17 \; 24 \; 28$ & $0.114$ & \phm{$<$}$\;-25.32$ & \phm{$<$}$\;30.16$ & \phm{$<$}$\;-29.67$ & \phm{$<$}$\;25.81$ & \phm{$<$}$\;-1.67$ \\
PG 1626+554 & $16 \; 27 \; 56.2$ & $+55 \; 22 \; 32$ & $0.133$ & \phm{$<$}$\;-25.87$ & \phm{$<$}$\;29.74$ & \phm{$<$}$\;-29.34$ & \phm{$<$}$\;26.27$ & \phm{$<$}$\;-1.33$ \\
PG 2130+099 & $21 \; 32 \; 27.82$ & $+10 \; 08 \; 19.2$ & $0.061$ & \phm{$<$}$\;-25.20$ & \phm{$<$}$\;29.72$ & \phm{$<$}$\;-28.96$ & \phm{$<$}$\;25.97$ & \phm{$<$}$\;-1.44$ \\
PG 2214+139 & $22 \; 17 \; 12.26$ & $+14 \; 14 \; 20.9$ & $0.067$ & \phm{$<$}$\;-25.44$ & \phm{$<$}$\;29.57$ & \phm{$<$}$\;-30.66$ & \phm{$<$}$\;24.35$ & \phm{$<$}$\;-2.00$ \\
PG 2233+134 & $22 \; 36 \; 07.68$ & $+13 \; 43 \; 55.3$ & $0.325$ & \phm{$<$}$\;-25.90$ & \phm{$<$}$\;30.52$ & \phm{$<$}$\;-30.03$ & \phm{$<$}$\;26.39$ & \phm{$<$}$\;-1.59$ 
  \enddata
\end{deluxetable*}

We interpolate the optical data to find the monochromatic luminosity
at 2500~\AA\ for each \mbox{COMBO-17} source using the redshifts and
the 12 medium-band photon fluxes given by \citet{wolf04}.  These bands
range between $4180-9140$~\AA\ which, for rest-frame 2500~\AA,
corresponds to a redshift range of $0.67<z<2.66$.  For the 15 sources
outside of this redshift range we linearly extrapolate to find the
2500~\AA\ luminosity.  We calculated the monochromatic 2~keV
luminosities from the $0.5-2$~keV luminosities, assuming a power-law
spectral energy distribution (SED) with $\Gamma=2.0$.
Table~\ref{c17_summary} lists the properties of the \mbox{COMBO-17}
AGN sample, including the \mbox{COMBO-17} ID from \citet{wolf04}, the
\mbox{E-CDF-S} ID from \citet{lehmer05b}, the sources' positions and
redshifts, monochromatic 2500~\AA\ and 2~keV fluxes and luminosities,
\alphaox , the effective \xray\ power-law photon index ($\Gamma$),
$R$, and a column identifying blended or extended \mbox{COMBO-17}
AGNs.

Since the majority of the \mbox{COMBO-17} AGNs do not have optical
spectra, we cannot easily identify obscured AGNs (using emission-line
and absorption-line diagnostics) that may contaminate our sample.  To
identify potential obscured \mbox{COMBO-17} AGNs, we will test the
\alphaox\ correlations below after removing the 22 sources with an
effective \xray\ power-law photon index $\Gamma<1.6$, which is
indicative of \xray\ obscuration \citep{lehmer05b}.  This procedure
does not help to identify potential obscured AGNs in the small number
of \mbox{COMBO-17} sources without \xray\ counterparts; these sources
will need future spectroscopic observations to identify possible
absorption.

The \alphaox\ distribution of our \mbox{COMBO-17} AGN sample is shown
in Figure~\ref{alphaox_hist}.  \mbox{COMBO-17} AGNs that have measured
effective photon indices of $\Gamma < 1.6$, and sources that have
\alphaox\ upper-limits, are shown as hatched and shaded histograms,
respectively.  The \mbox{COMBO-17} AGNs typically have flatter
\alphaox\ values than those calculated for more luminous AGN samples.
This is consistent with the observed anti-correlation between
\alphaox\ and optical/UV luminosity seen in the studies mentioned in
Section~1, where the value of \alphaox\ falls with increasing \luv.
For comparison, we mark the mean \alphaox\ values for the three AGN
samples used in S05 in Figure~\ref{alphaox_hist}.

\subsubsection{Bright Quasar Survey}

To increase the luminosity range covered at low redshifts we included
a set of 46 BQS quasars with $z<0.5$ analyzed by \citet*{brandt00}.
While the BQS sample does extend beyond \mbox{$z=0.5$},
\citet{jester05a} found a bias against detecting AGNs in the BQS
catalog in the redshift range $0.5<z<1.0$.  Since the BQS sample is
meant to be a higher-luminosity complement to the S05 and Seyfert 1
samples at low redshifts, we removed 21 AGNs with $M_B > -23$.  This
luminosity threshold also removes fainter AGNs that are more likely to
be affected by contamination from their host galaxies.  We excluded 13
radio-loud AGNs and five known BALQSOs \citep[see footnote 4
in][]{brandt00}.  In addition, following the arguments of
\citet{brandt00}, we also exclude two BQS quasars that have
significant UV absorption, but do not meet the formal criteria for
BALQSOs: PG~$1351+640$ \citep[e.g.,][]{zheng01} and PG~$1411+442$
\citep[e.g.,][]{malkan87}.

We converted the $f_{\mbox{\scriptsize 3000 \AA}}$ values used in
\citet{brandt00} to \luv , assuming $f(\nu) \propto \nu^{-\alpha}$
with $\alpha = -0.5$.  We used PIMMS\footnotemark\
\footnotetext{Portable Interactive Multi-Mission Simulator available
  at {\ttfamily
      http://heasarc.gsfc.nasa.gov/\discretionary{}{}{}Tools/\discretionary{}{}{}w3pimms.html}}
to calculate the monochromatic 2~keV luminosity, assuming a power-law
spectrum with $\Gamma = 2$, from the \rosat\ \mbox{$0.5-2.0$~keV}
pointed PSPC count-rate when available, and from the
\mbox{$0.5-2.0$~keV} {\slshape ROSAT} All Sky Survey (RASS) count-rate
when no pointed \rosat\ observation existed.  An \xray\ upper limit
for one BQS quasar with no \xray\ counterpart was calculated from the
RASS.  Our BQS sample contains 46 sources, with 45 ($98\%)$ having
\xray\ detections.  We use the astrometry presented in
\citet{jester05a} for the BQS sources included therein and the
astrometry from \citet{veron-cetty03} for the remaining sources.
Table~\ref{bqs_summary} presents the BQS sources' names, positions,
and redshifts along with the monochromatic 2500 \AA\ and 2~keV fluxes
and luminosities, and \alphaox .

\subsubsection{Additional High-z AGNs}

%
%
\begin{figure*}
\epsscale{1.15}
\plottwo{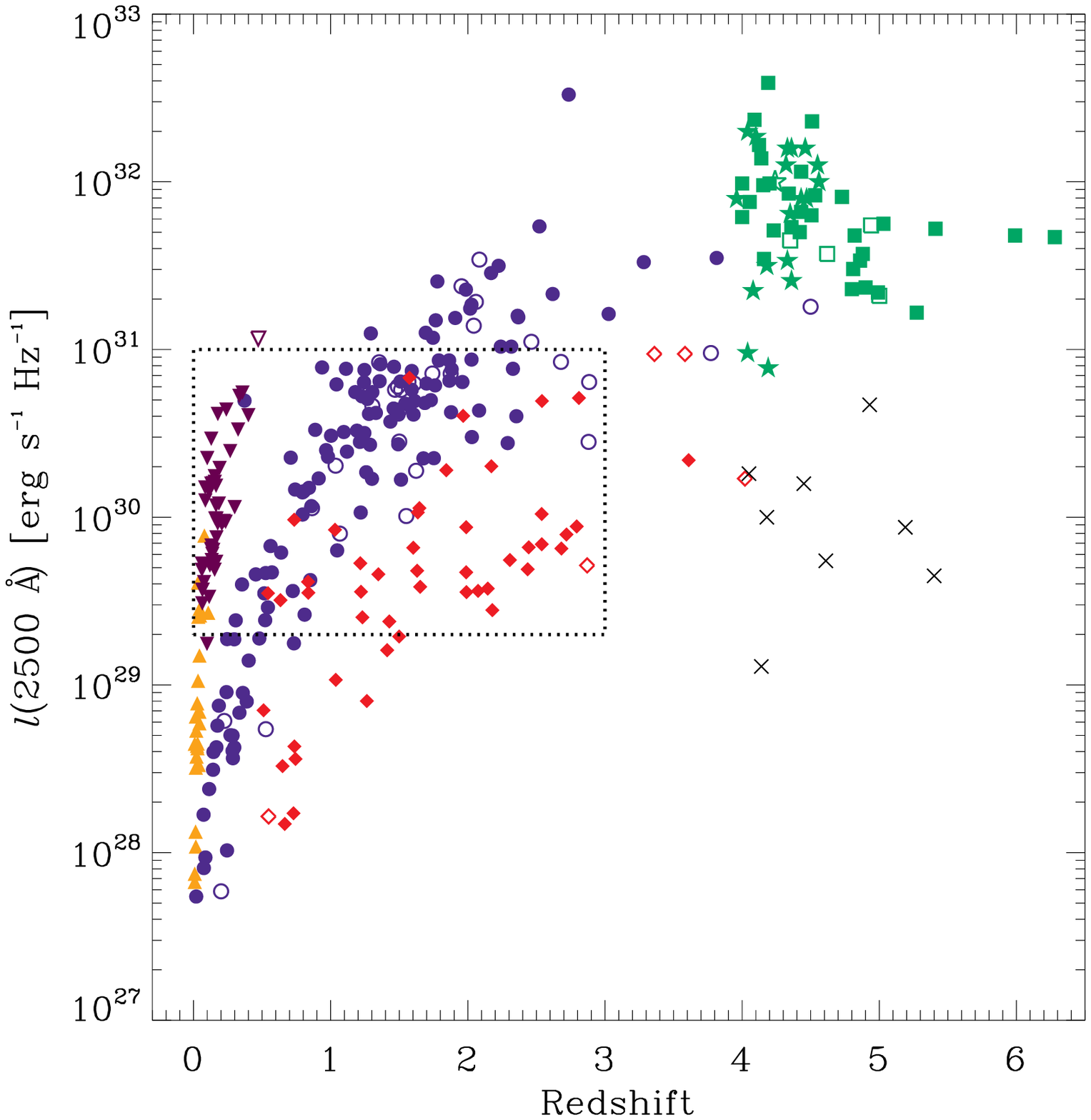}{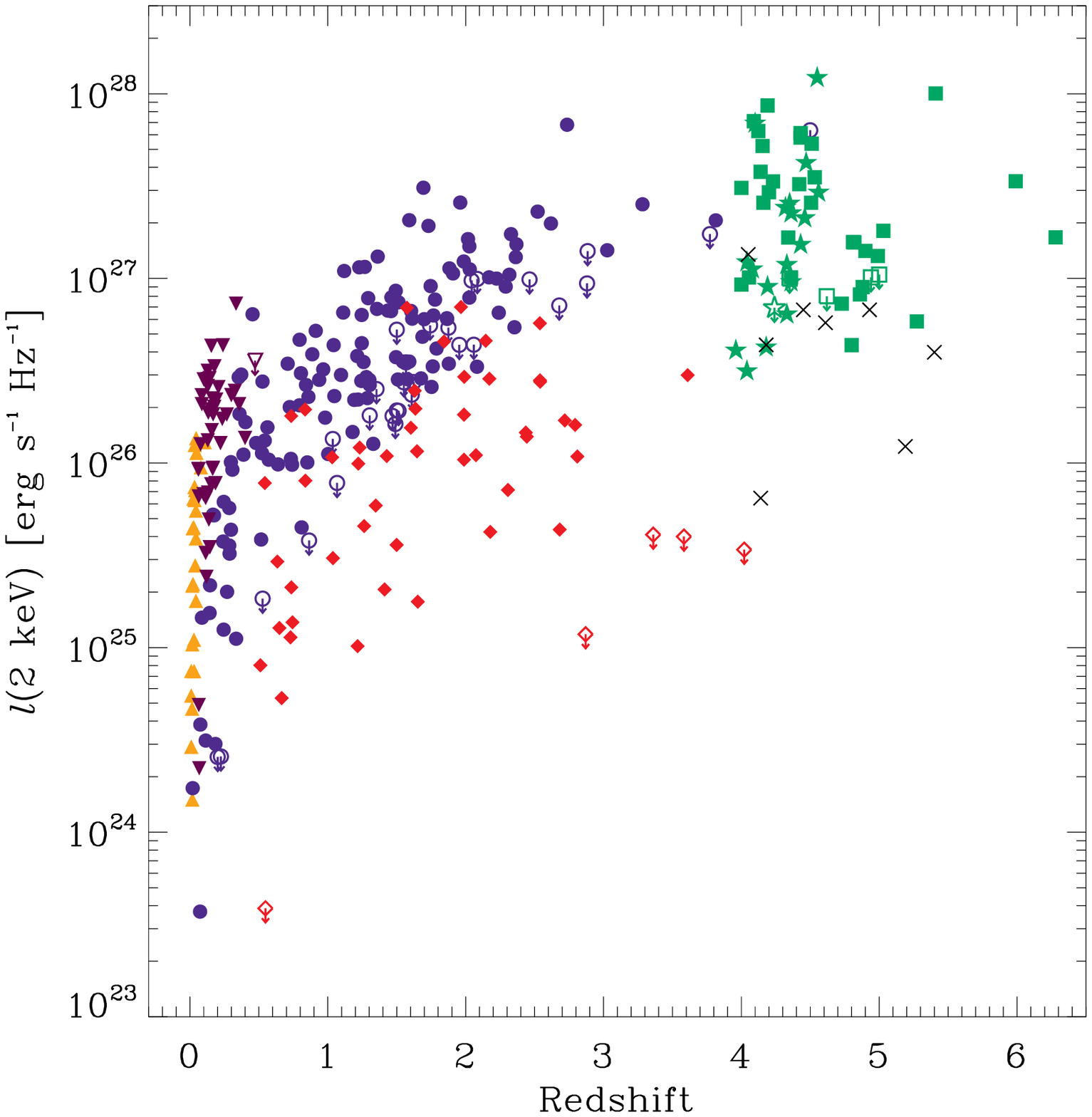}
\caption{\label{lum} 2500~\AA\ luminosity ({\slshape left}) and 2~keV
  luminosity ({\slshape right}) vs. redshift for the S05 Seyfert~1
  ({\slshape orange upward-pointing triangles}), SDSS ({\slshape blue circles}),
  and high-redshift ({\slshape green boxes}) samples shown with the
  \mbox{COMBO-17} ({\slshape red diamonds}), BQS ({\slshape
    violet downward-pointing triangles}), and optically selected,
  high-redshift AGNs ({\slshape green stars}).  \xray\ upper limits are
  represented using open symbols ({\slshape left}) or arrows
  ({\slshape right}).  The \luv\ - $z$ subsample of 187 sources used
  in \S 3.6 is denoted by the dotted-line box ({\slshape left}). The
  properties of the known \xray --selected AGNs at $z>4$ ({\slshape
    black crosses}) are also shown; these sources are not included in our
  formal analyses due to their different selection criteria.}
\end{figure*}

To increase the size of our high-redshift sample we included recently
published, optically-selected AGNs with $z>4$.  We included the ten
luminous quasars listed in Table~3 of \citet{vignali05a} (excluding
the BALQSO PSS~$1506+5220$), and three radio-quiet AGNs from
\citet{shemmer05a} that were not included in S05 (Q~$0000-263$,
BR~$0351-1034$, and BR~$2237-0607$). To extend the coverage of our
sample to somewhat lower luminosities at $z>4$ we included a sample of
six optically-selected, faint, radio-quiet, non-BALQSO sources
targeted in \mbox{$>10$~ks} \chandra\ observations \citep{kelly05}.
We analyzed these \xray\ and optical data in the same manner as
outlined in \citet{vignali05a}.  We briefly present these data in the
Appendix.  The addition of these three optically-selected, $z>4$ AGN
samples adds 19 optically-selected AGNs to our total sample, 18
($94\%)$ of which have \xray\ detections.

For comparison purposes, we also plot the eight\footnotemark\ published
\xray--selected, radio-quiet, \mbox{$z>4$}, AGNs ({\slshape crosses}) in
Figures~3$-$9.  We present the observed properties of these sources in
Table~\ref{xrs} in the Appendix.  While these \xray --selected AGNs substantially
extend the coverage of the UV luminosity - redshift plane at high
redshifts (see Figure~\ref{lum}a), we do not include these AGNs in our
formal analyses as they have been selected with different criteria than the
other optically-selected AGN samples.  However, in \S4.3 we discuss
how these AGNs fit into the relations derived in \S3 using the
optically-selected AGN samples.

\footnotetext{We excluded the two known \xray--selected, radio-loud,
  $z>4$ AGNs, RX J1028.6-0844 \citep[$z=4.28$;][]{zickgraf97} and RX
  J1759.4$+$6638 \citep[$z=4.32$;][]{henry94}.  An up-to-date
  compilation of $z>4$ AGNs with \xray\ detections is available at
  {\ttfamily
      http://www.astro.psu.edu/\discretionary{}{}{}users/\discretionary{}{}{}niel/\discretionary{}{}{}papers/\discretionary{}{}{}highz-xray-detected.dat}}

\pagebreak
\subsection{Full Sample}

\begin{deluxetable}{lcccc}[b]
  \tablewidth{0pt}
  \tablecolumns{5}
  \tablecaption{\label{sample_numbers}Summary of Samples Utilized}
  \tablehead{
    \colhead{Sample} &
    \colhead{Total} &
    \colhead{Number} & 
    \colhead{$0.5-2.0$ keV Limit} & 
    \colhead{Area} \\
    \colhead{} &
    \colhead{AGNs} &
    \colhead{\xray\ Detected} &
    \colhead{[erg cm$^{-2}$ s$^{-1}$]} &
    \colhead{[sq. deg.]} 
  }
  \startdata
    \cutinhead{S05}
      SDSS & 155 & $126 \; $\phn$(81\%)$ &  $\sim10^{-14}$ & $15$ \\
      Seyfert 1 & \phm{\tablenotemark{a}}\phn25\tablenotemark{a} & \phn$25 \; (100\%)$ & $\sim10^{-12}$ & \nodata\tablenotemark{b} \\
      High-$z$ & \phn36 & \phn$32 \; $\phn$(89\%)$ & $\sim10^{-15}$ & \nodata\tablenotemark{b} \\ 
    \cutinhead{This work}
      COMBO-17 & \phn52 & \phn$47 \; $\phn$(90\%)$ & $\sim10^{-16}$ & $0.26$ \\
      BQS & \phn46 & \phn$45 \; $\phn$(98\%)$ & $\sim10^{-13}$ & $10,714$ \\
      High-$z$ & \phn19 & \phn$18 \; $\phn$(95\%)$ & $\sim10^{-15}$ & \nodata\tablenotemark{b} \\
    \cutinhead{Full Sample}
      Total & 333 & $293 \; $\phn$(88\%)$ & \nodata & \nodata\tablenotemark{b} 
  \enddata
\tablenotetext{a}{After removing the twelve sources that overlap with the BQS sample.}
\tablenotetext{b}{Not well defined.}
\end{deluxetable}

The full sample used in our analysis consists of 333 AGNs, 293
($88\%$) of which have \xray\ detections; this is a much higher \xray\
detection fraction than for most previous analyses, where the
detection fraction is typically $10-50\%$.  The number of AGNs
contributed by each sample, and their respective \xray\ detection
fractions, are summarized in Table~\ref{sample_numbers}.
Figure~\ref{lum} shows the monochromatic 2500~\AA\ ({\slshape left})
and 2~keV ({\slshape right}) luminosity versus redshift for the
combined AGN sample.  The addition of the \mbox{COMBO-17} ({\slshape
  diamonds}) and BQS ({\slshape downward-pointing triangles}) AGNs to
the S05 data substantially improves the luminosity range covered out
to $z\sim3$, corresponding to $\simeq 85\%$ of cosmic history.

\section{Partial Correlation and Linear Regression Analyses}
To measure properly the correlations among $l_{\mbox{\scriptsize 2500
    \AA}}$, $l_{\mbox{\scriptsize 2~keV}}$, \alphaox , and redshift,
we must attempt to eliminate (or minimize) potential sources of bias
from our analysis.  Following the arguments of \citet*{kembhavi86}, we
use the AGN luminosities in our calculations and not the measured
fluxes which, if used, can yield different results depending upon the
cosmology and $k$-corrections assumed.  However, using the
luminosities can introduce a different bias since luminosity and
redshift are typically correlated in flux-limited samples.  Our sample
was constructed to cover a large area of the $l - z$ plane, which acts
to break this degeneracy.

While the majority of the AGNs in our sample have \xray\ detections
($88\%$), we cannot overlook the effects of sources with \xray\ upper
limits.  In addition, our sample data undoubtedly contain intrinsic
scatter that is not quantified in our measurements.  We have minimized
the effects of some sources of intrinsic scatter such as host-galaxy
contamination, obscuration, and excess UV and \xray\ flux associated
with radio jets (as discussed in \S2.2.1).  Another source of
intrinsic scatter in our data is caused by AGN variability, which
affects our calculations of \alphaox\ since the optical and \xray\
observations were not taken concurrently.  Unfortunately, the effects
of variability cannot be corrected and remain a significant source of
intrinsic scatter in our data.  To measure the aforementioned
correlations among these AGNs, proper statistical tools must be used
that take these issues into consideration.
\pagebreak
\subsection{Statistical Tools}

%
%
\begin{figure*}
\epsscale{1.0}
\plotone{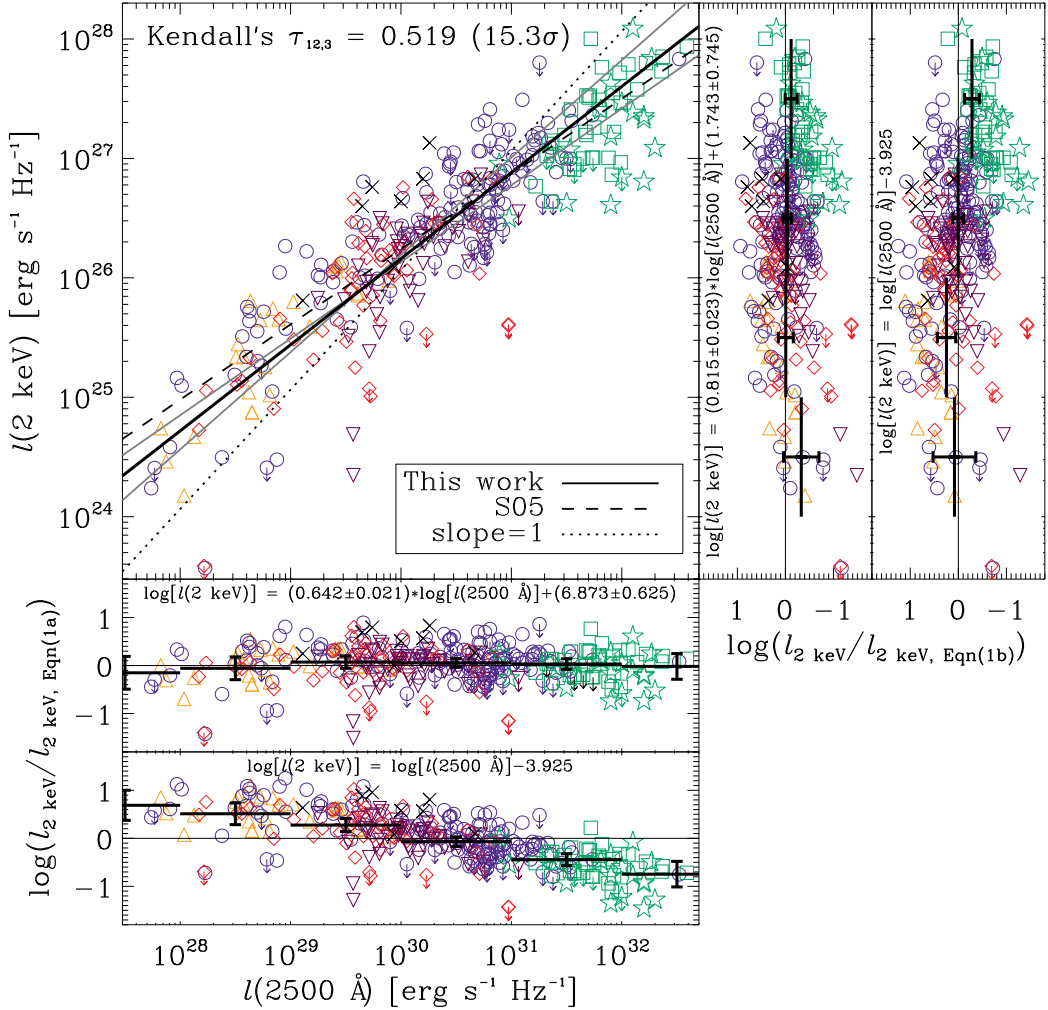} 
\caption{\label{uv_vs_x}
 ({\slshape top}) Rest-frame 2~keV monochromatic luminosity versus
 rest-frame 2500~\AA\ monochromatic luminosity.  The symmetric,
 best-fit \lx - \luv\ relationship, given by
 Equation~(\ref{eqn_luv_lx_3}), is denoted by a solid, black line.
 Equations~(\ref{eqn_luv_lx_1}) and (\ref{eqn_luv_lx_2}) are denoted
 by solid, gray lines in order of increasing \lx\ at \luv$ = 10^{32}
 \mbox{ erg s}^{-1}~\mbox{Hz}^{-1}$.  For comparison, the best-fit
 line derived by S05 ({\slshape dashed line}) and a $\beta = 1$
 relation ({\slshape dotted line}), normalized to best fit the data,
 are shown.  The residuals for Equation~(\ref{eqn_luv_lx_1}) and the
 $\beta=1$ relation are shown in the two lower panels,
 respectively. The residuals for Equation~(\ref{eqn_luv_lx_2}) and the
 $\beta=1$ relation are shown in the two panels on the right,
 respectively.  The overlaid error bars denote the mean and the
 $3\sigma$ standard deviation of the mean of the residuals calculated
 for each $\Delta {\rm log}(l_{\mbox{\scriptsize 2500 \AA}})=1$ bin.
 Symbols are defined as in Figure~\ref{lum}, although all symbols are
 plotted as open to minimize symbol crowding. Limits are denoted with
 arrows.  We find a highly significant ($15.3\sigma$) correlation
 between \lx\ and \luv\ (controlling for redshift) with a best-fit
 slope of $\beta=0.73\pm0.1$, calculated from the ILS bisector.}
\end{figure*}

While our extended coverage of the $l-z$ plane helps to break the
redshift bias introduced by using luminosities instead of fluxes in
our analysis, a correlation between $l$ and $z$ can still be clearly
seen in Figure~\ref{lum}.  To take into account this existing
correlation we use a partial-correlation analysis method that is
designed to measure the correlation between two variables, controlling
for the effects of one or more additional variables.  A
partial-correlation method that properly handles censored data was
developed by \citet{akritas96a}.  This method builds upon Kendall's
rank-correlation coefficient \citep{kendall38} and Kendall's partial
rank-correlation coefficient \citep{kendall70} to include censored
data.  To measure the partial correlations we use the FORTRAN program
CENS\_TAU, available from the Penn State Center for
Astrostatistics\footnotemark, which uses the methodology presented in
\citet{akritas96a}.  We present the results of our partial-correlation
analysis in Table~\ref{correlations}.  In this table we also present
the significance of the correlations with the potentially obscured
($\Gamma < 1.6$, see \S~2.2.1) \mbox{COMBO-17} AGNs removed.  With the
exception of a slight decrease in the significance in the \lx - \luv\
correlation, there is no substantial change in the results when these
sources are removed.  Therefore, we use the full sample in our
subsequent analyses.

\footnotetext{Available at {\ttfamily
    http://www.astrostatistics.psu.edu/\discretionary{}{}{}statcodes/\discretionary{}{}{}cens\us tau}}

To derive the linear-regression parameters for correlations, we use
the Astronomy SURVival Analysis software package \citep[ASURV Rev
1.2;][]{isobe90,lavalley92}, which implements the bivariate
data-analysis methods presented in \citet*{isobe86}.  ASURV also
properly handles censored data using the survival-analysis methods
presented in \citet{feigelson85} and \citet{isobe86}.  We performed
linear regressions using both the fully parametric EM (estimate and
maximize) regression algorithm \citep*{dempster77} and the
semi-parametric Buckley-James regression algorithm \citep{buckley79}.
The EM Regression algorithm uses an iterative least squares (ILS)
method that reduces to the traditional ordinary least squares (OLS)
when no censored data are present.  We present the parameters provided
by the EM regression algorithm below, but in all cases the results
from the Buckley-James regression algorithm agreed within the errors.

\boldmath
\subsection{$l_{\mbox{\scriptsize 2 keV}}$ versus $l_{\mbox{\scriptsize 2500 \AA}}$}
\unboldmath

Early \xray\ studies of AGNs revealed a correlation between \xray\ and
UV monochromatic luminosities, typically found to be
$l_{\mbox{\scriptsize X}} \propto l_{\mbox{\scriptsize UV}}^{\beta}$,
where $\beta \simeq 0.7-0.8$ (see the references in \S1).  The
observational evidence that $\beta$ is not unity was unexpected and
led to suggestions that this finding was not an intrinsic property of
AGNs, but rather an observational bias introduced as a result of the
larger scatter of $l_{\mbox{\scriptsize UV}}$ compared to
$l_{\mbox{\scriptsize X}}$
\citep[e.g.,][]{chanan83,yuan98}. \citet{la_franca95} found an
$l_{\mbox{\scriptsize X}} - l_{\mbox{\scriptsize UV}}$ correlation
consistent with $\beta = 1$ using a generalized orthogonal regression
procedure that takes into account measurement errors on both variables
and intrinsic scatter.  However, this method does not properly take
into account the effects of censored data, which typically affect
optically-selected AGN samples.  Currently there is no statistical
analysis method that properly handles data with both censoring and
measurement errors (M.~Akritas 2005, private communication).

For our optically-selected AGN sample we find a highly significant
($15.3\sigma$) correlation between the observed monochromatic \xray\
and UV luminosities.  The EM linear-regression algorithm in ASURV that
we use to calculate the linear-regression parameters is based on the
traditional ordinary least-squares method, OLS($Y \mid X$), which
minimizes the residuals of the dependent variable, $Y$ (the EM method
assumes the residuals are Gaussian distributed).  Since both \lx\ and
\luv\ are observed, neither can be truly called the ``independent'' or
``dependent'' variable.  Indeed, a different result can be obtained if
the residuals of the independent variable are instead minimized [i.e.,
OLS($X \mid Y$)].  Rather than use the traditional OLS($Y \mid X$), we
choose a method that instead provides a symmetric fit to the data.
Following the arguments in \S~5 of \citet{isobe90}, we use the OLS
bisector, which simply bisects the two lines from the OLS($Y \mid X$)
and OLS($X \mid Y$), or in our case ILS($Y \mid X$) and ILS($X \mid
Y$) from the EM Algorithm in ASURV.

We perform linear regressions with ASURV on the full sample of 333
AGNs and find the relation between \lx\ and \luv\ to be
\begin{subequations}
\begin{equation}
\label{eqn_luv_lx_1}
{\rm log}(l_{\mbox{\scriptsize 2 keV}}) = (0.64 \pm 0.02) \; {\rm log}(l_{\mbox{\scriptsize 2500 \AA}}) + (6.87 \pm 0.63)
\end{equation}
 using ILS($Y \mid X$) (i.e., treating \lx\ as the dependent variable) and
\begin{equation}
\label{eqn_luv_lx_2}
{\rm log}(l_{\mbox{\scriptsize 2 keV}}) = (0.82 \pm 0.02) \; {\rm log}(l_{\mbox{\scriptsize 2500 \AA}}) + (1.74 \pm 0.75)
\end{equation}
using ILS($X \mid Y$) (i.e., treating \luv\ as the dependent
variable).  Note that the slope of the \lx - \luv\ relation does
depend on which luminosity is used as the dependent variable, but both
cases are inconsistent with $\beta = 1$.  We use the equations given
in Table~1 of \citet{isobe90} to calculate the bisector of the two
regression lines.  We find the ILS bisector to be
\begin{equation}
\label{eqn_luv_lx_3}
{\rm log}(l_{\mbox{\scriptsize 2 keV}}) = (0.72 \pm 0.01) \; {\rm log}(l_{\mbox{\scriptsize 2500 \AA}}) + (4.53 \pm 0.69)
\end{equation}
\end{subequations}

In Figure~\ref{uv_vs_x}, we show \lx\ versus \luv\ for the full AGN
sample ({\slshape top panel}).  We show the symmetric, best-fit \lx -
\luv\ relationship, given by Equation~(\ref{eqn_luv_lx_3}), as a
solid, black line.  Equations~(\ref{eqn_luv_lx_1}) and
(\ref{eqn_luv_lx_2}) are denoted by solid, gray lines in order of
increasing \lx\ at \luv$ = 10^{32} \mbox{ erg s}^{-1}~\mbox{Hz}^{-1}$.
For comparison purposes we show the \lx - \luv\ relation found by S05
({\slshape dashed line}), along with a $\beta = 1$ relation with a
normalization chosen to minimize the \lx\ residuals ({\slshape dotted
  line}).  The residuals for Equation~(\ref{eqn_luv_lx_1}) and the fit
assuming $\beta=1$ are given in the bottom panels, with the mean and
the ($3\sigma$) standard deviation of the mean calculated for sources
in each $\Delta {\rm log}(l_{\mbox{\scriptsize 2500 \AA}})=1$ bin
denoted with large error bars.  The residuals for
Equation~(\ref{eqn_luv_lx_2}) and the fit assuming $\beta=1$ are given
in the right-hand panels using the same symbols.  From the systematic
$\beta=1$ residuals present along both axes, it is apparent that a
$\beta=1$ relation provides an unsatisfactory fit to the data,
independent of the luminosity considered as the dependent variable.

\boldmath
\subsection{\alphaox\ versus $l_{\mbox{\scriptsize 2500 \AA}}$}
\unboldmath
%
%
\begin{figure}
\epsscale{1.1}
\plotone{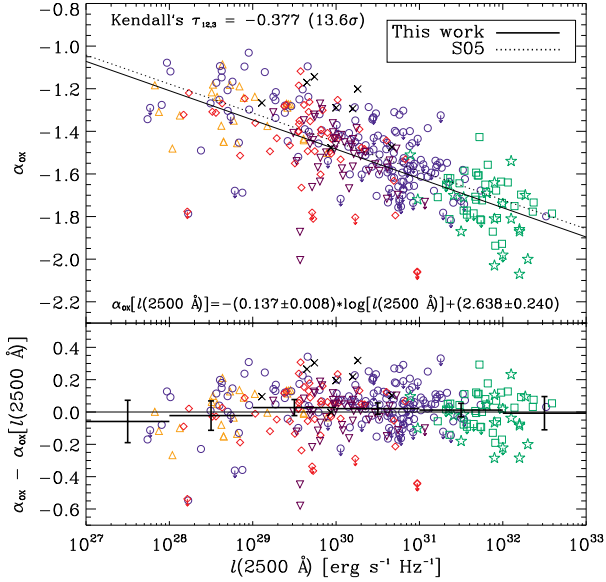}
\caption{\label{luv_vs_aox}
 ({\slshape top}) \alphaox\ vs. rest-frame 2500~\AA\ monochromatic
 luminosity.  The best-fit line from Equation~(\ref{eqn_aox_luv}) is
 shown ({\slshape solid line}).  For comparison, the best fit line
 derived by S05 ({\slshape dotted line}) is also shown. ({\slshape
 bottom}) Residuals from the fit shown in the top panel.  The overlaid
 error bars denote the mean and the $3\sigma$ standard deviation of
 the mean of the residuals calculated for each $\Delta {\rm
 log}(l_{\mbox{\scriptsize 2500 \AA}})=1$ bin.  Symbols are defined as
 in Figure~\ref{lum}, although all symbols are plotted as open to
 minimize symbol crowding. Limits are denoted with arrows.  We find a
 highly significant ($13.6\sigma$) correlation between \alphaox\ and
 \luv\ (controlling for redshift).}
\end{figure}

Given the highly significant correlation between \lx\ and \luv\ found
for our AGN sample, we investigate how \alphaox\ changes with respect
to \luv , \lx, and redshift.  For our optically-selected AGN sample we
confirm a highly significant ($13.6\sigma$) anti-correlation between
\alphaox\ and \luv\ when controlling for the effects of redshift.  The
best-fit parameters for the \alphaox - \luv\ relation are
\begin{equation}
\label{eqn_aox_luv}
\alpha_{\mbox{\tiny OX}} = (-0.137 \pm 0.008) \; {\rm log}(l_{\mbox{\scriptsize 2500 \AA}}) + (2.638 \pm 0.240)
\end{equation}

In Figure~\ref{luv_vs_aox}, we show \alphaox\ versus
$l_{\mbox{\scriptsize 2500 \AA}}$ for our full AGN sample.  We show
the best-fit linear regression found for this sample, given in
Equation~(\ref{eqn_aox_luv}), as a solid line.  The residuals for the
fit are given in the bottom panel, with the mean and standard
deviation of the mean calculated for sources in each $\Delta {\rm
  log}(l_{\mbox{\scriptsize 2500 \AA}})=1$ bin denoted with $3\sigma$
error bars.  For comparison purposes we show the \alphaox - \luv\
relation found by S05 ({\slshape dotted line}).

\boldmath
\subsection{\alphaox\ versus $l_{\mbox{\scriptsize 2 keV}}$}
\unboldmath

%
%
\begin{figure}
\epsscale{1.1}
\plotone{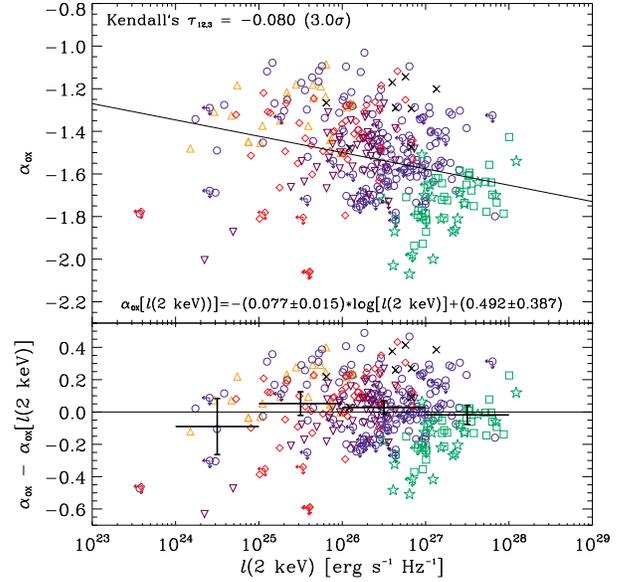}
\caption{\label{lx_vs_aox}
 ({\slshape top}) \alphaox\ vs. rest-frame 2~keV monochromatic
 luminosity.  The best-fit line from Equation~(\ref{eqn_aox_lx}) is
 shown ({\slshape solid line}).  ({\slshape bottom}) Residuals from
 the fit shown in the top panel.  The overlaid error bars denote the
 mean and the $3\sigma$ standard deviation of the mean of the
 residuals calculated for each $\Delta {\rm log}(l_{\mbox{\scriptsize
 2 keV}})=1$ bin.  Symbols are defined as in Figure~\ref{lum},
 although all symbols are plotted as open to minimize symbol
 crowding. Limits are denoted with arrows.  We find a lower, but still
 significant ($3.0\sigma$) correlation between \alphaox\ and \lx\
 (controlling for redshift).}
\end{figure}

The correlation between \alphaox\ and \lx\ has been previously
examined \citep{green95}, but the AGN samples used had low \xray\
detection fractions and included both radio-loud AGNs and BALQSOs,
both of which can obfuscate the intrinsic \xray\ emission of interest
here (see \S~1).  In addition, \citet{green95} did not check if the
correlation between \alphaox\ and \lx\ was the result of a true
correlation, or a by-product of the known correlation between \lx\ and
redshift.

The high \xray\ detection fraction of our AGN sample allows us to
examine the correlation between \alphaox\ and \lx .  We find a weaker,
but still significant ($3.0\sigma$) anti-correlation between \alphaox\
and \lx\ when controlling for the effects of redshift and taking into
account the double-censoring present in this relation (i.e., censoring
of both the dependent and independent variables).  The
linear-regression methods used previously to derive the parameters of
the correlations (EM and Buckley-James regression algorithms) are only
strictly valid if double-censoring is not present.  However, given the
high \xray\ detection fraction of our sample, we continue to use the
aforementioned EM and Buckley-James regression methods, assuming the
independent variable (\lx) is detected.  Note that our choice of
linear-regression techniques does not affect the significance of the
anti-correlation presented above.

The best-fit parameters for the \alphaox - \lx\ relation from the EM
regression method are
\begin{equation}
\label{eqn_aox_lx}
\alpha_{\mbox{\tiny OX}} = (-0.077 \pm 0.015) \; {\rm log}(l_{\mbox{\scriptsize 2 keV}}) + (0.492 \pm 0.387)
\end{equation}

In Figure~\ref{lx_vs_aox}, we show \alphaox\ versus \lx\ for our full
AGN sample; the best-fit linear regression found for this sample,
given in Equation~(\ref{eqn_aox_lx}), is shown by a solid line.  The
residuals for the fit are given in the bottom panel, with the mean and
standard deviation of the mean calculated for sources in each $\Delta
{\rm log}(l_{\mbox{\scriptsize 2~keV}})=1$ bin denoted with $3\sigma$
error bars.

The clear trend of higher (i.e., less-negative) \alphaox\ values with
decreasing \luv\ seen in Fig~\ref{luv_vs_aox} is not as apparent in
the \alphaox - \lx\ plot.  This could be due, in part, to the smaller
range in \xray\ luminosity covered by the AGNs ($\Delta {\rm
  log}(l_{\mbox{\scriptsize 2~keV}})=4$) compared to the UV luminosity
range ($\Delta {\rm log}(l_{\mbox{\scriptsize 2500~\AA}})=5$).  In
addition, the scatter in the residuals for the \alphaox - \lx\
relation is larger than that seen for \luv .

\boldmath
\subsection{\alphaox\ versus redshift}
\unboldmath

%
%
\begin{figure}
\epsscale{1.1}
\plotone{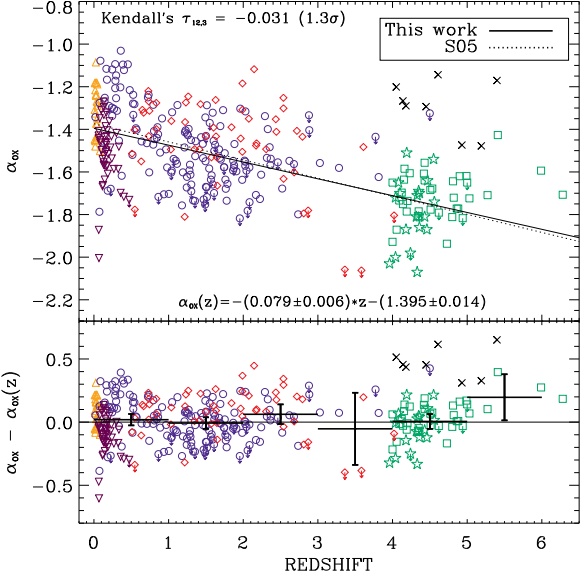}
\caption{\label{z_vs_aox}
 ({\slshape top}) \alphaox\ vs. redshift.  The best-fit line from
 Equation~(\ref{eqn_aox_z}) is shown ({\slshape solid line}).  For
 comparison, the best-fit line derived by S05 ({\slshape dotted line})
 is also shown. ({\slshape bottom}) Residuals from the fit shown in
 the top panel.  The overlaid error bars denote the mean and the
 $3\sigma$ standard deviation of the mean of the residuals calculated
 for each $\Delta z=1$ bin.  Symbols are defined as in
 Figure~\ref{lum}, although all symbols are plotted as open to
 minimize symbol crowding. Limits are denoted with arrows.  We find no
 significant ($1.3\sigma$) correlation between \alphaox\ and redshift
 (controlling for \luv ).}
\end{figure}

To test for a possible redshift dependence of \alphaox\ we measure the
partial correlation of \alphaox\ and $z$, controlling for \luv .  We
do not find a significant correlation between \alphaox\ and redshift
($1.3\sigma$), consistent with previous \alphaox\ studies.  The
best-fit parameters for the \alphaox $-z$ relation are
\begin{equation}
\label{eqn_aox_z}
\alpha_{\mbox{\tiny OX}} = (-0.079 \pm 0.006) \; z \: - (1.395 \pm 0.014)
\end{equation}

In Figure~\ref{z_vs_aox}, we show \alphaox\ versus $z$ for our full
AGN sample.  We show the best-fit linear regression found for this
sample, given in Equation~(\ref{eqn_aox_z}), as a solid line.  The
apparent correlation between \alphaox\ and redshift is simply an
artifact of the \luv - $z$ correlation.  The residuals for the fit are
given in the bottom panel, with the mean and standard deviation of the
mean calculated for sources in each $\Delta z=1$ bin denoted with
$3\sigma$ error bars.  For comparison purposes we show the \alphaox
$-z$ relation found by S05 ({\slshape dotted line}).

%
%
\begin{figure}[b]
\epsscale{1.1}
\plotone{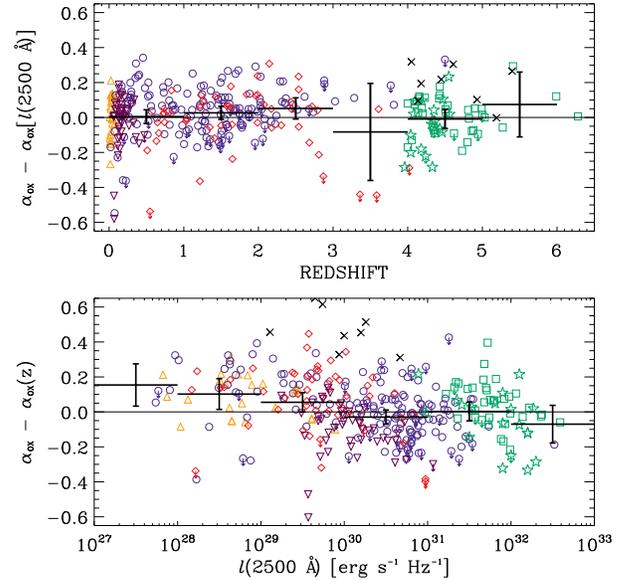}
\caption{\label{residuals}
 \alphaox\ residuals as a function of redshift ({\slshape top panel})
 and \luv\ ({\slshape bottom panel}).  The overlaid error bars denote
 the mean and the $3\sigma$ standard deviation of the mean of the
 residuals calculated for each $\Delta z=1$ bin ({\slshape top panel})
 or $\Delta {\rm log}(l_{\mbox{\scriptsize 2500 \AA}})=1$ bin
 ({\slshape bottom panel}).  Symbols are defined as in
 Figure~\ref{lum}, although all symbols are plotted as open to
 minimize symbol crowding. Limits are denoted with arrows.  The
 systematic residuals in the lower plot indicate that \alphaox\ cannot
 be dependent on redshift alone.}
\end{figure}

In Figure~\ref{residuals}, we show the \alphaox $-$ \alphaox$($\luv$)$
residuals as a function of redshift ({\slshape top panel}) and the
\alphaox $-$ \alphaox$(z)$ residuals as a function of \luv\ ({\slshape
  bottom panel}).  It is clear from Figure~\ref{residuals} that there
is a strong luminosity dependence in the \alphaox $-$ \alphaox$(z)$
residuals, while the \alphaox $-$ \alphaox$($\luv$)$ residuals show no
strong redshift dependence.  Using ASURV to find \alphaox\ as a
function of {\slshape both} \luv\ and redshift, we find the following
relation.
\begin{eqnarray}
\alpha_{\mbox{\tiny OX}}  & = & (-0.126 \pm 0.013) \; {\rm log}(l_{\mbox{\scriptsize 2500 \AA}}) \nonumber \\
 & & - (0.010 \pm 0.009) \; z + (2.311 \pm 0.372) \label{eqn_aox_luv_z}
\end{eqnarray}
Equation~(\ref{eqn_aox_luv_z}) shows that the coefficient of $z$ is
statistically consistent with zero ($1.1\sigma$), which agrees with
the findings of S05 and is consistent with \alphaox\ being independent
of redshift.

We use the \alphaox\ $-$ \alphaox $($\luv $)$ residuals to constrain
the maximum possible residual dependence of \alphaox\ on redshift,
which constrains the maximum amount the ratio of the UV-to-\xray\
luminosity can evolve with redshift.  We find the best-fit parameters
of the residuals to be $<$\alphaox\ $-$ \alphaox $($\luv $)> = (-0.004
\pm 0.006) \:z - (0.004 \pm 0.012)$.  The maximum absolute $1\sigma$
value for the slope of the residuals is $0.010 \;z$, which at $z=5$ is
equal to $0.05$.  The ratio between the UV and \xray\ flux is, by
definition, $r=f_{\nu}(\mbox{2500 \AA}) / f_{\nu}(\mbox{2 keV}) =
10^{2.606\alpha_{\mbox{\tiny OX}}}$.  Differentiating with respect to
\alphaox , we get $\delta r/r = 2.606 \; \mbox{ln}(10) \;
\delta\alpha_{\mbox{\tiny OX}} \simeq 6 \,\delta\alpha_{\mbox{\tiny
    OX}} \simeq 0.30$, for $\delta\alpha_{\mbox{\tiny OX}} = 0.05$.
Thus the the ratio of UV-to-\xray\ flux has not changed by more than
$30\% (1\sigma)$ over the redshift range $z=0-5$.

%
%
\begin{figure}
\epsscale{1.2}
\plotone{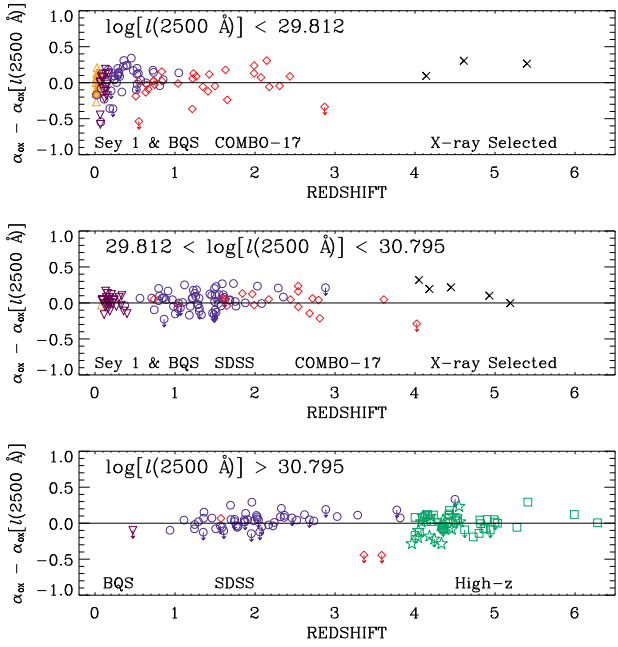}
\caption{\label{residuals_lum_breakdown}
 \alphaox\ residuals as a function of redshift shown for three
 luminosity ranges. Symbols are defined as in Figure~\ref{lum},
 although all symbols are plotted as open to minimize symbol
 crowding. Limits are denoted with arrows.  There is no apparent redshift
 dependence of the \alphaox\ $-$ \alphaox $($\luv $)$ residuals within
 any luminosity range.}
\end{figure}

To examine if \alphaox\ is independent of redshift over the range of
luminosities observed, we divided the AGNs equally into three
luminosity bins, with each bin containing 111 sources.  We plot the
\alphaox\ $-$ \alphaox $($\luv $)$ residuals as a function of redshift
for each UV luminosity bin in Figure~\ref{residuals_lum_breakdown}.
It is apparent that the \mbox{COMBO-17} sample and the Seyfert~1
sample of S05 comprise most of the lower-luminosity AGNs (${\rm
  log}(l_{\mbox{\scriptsize 2500 \AA}})<29.812$) .  The number of
\mbox{COMBO-17} AGNs decreases in the moderate UV luminosity bin where
the BQS and SDSS AGNs become prevalent ($27.812 < {\rm
  log}(l_{\mbox{\scriptsize 2500 \AA}})<30.795$).  At the highest
luminosities, the $z>4$ AGNs and the SDSS sample dominate (${\rm
  log}(l_{\mbox{\scriptsize 2500 \AA}})>30.795$).  None of the
residuals shows any indications of having a systematic offset, showing
that AGNs spanning five decades in UV luminosity show little \alphaox\
evolution with redshift.

\boldmath
\subsection{Comparing \alphaox\ - $z$ and \lx\ - $z$ relations}
\unboldmath

Recent discussions have questioned the validity of measuring the
redshift evolution of \alphaox\ controlling for the effects of \luv ,
since \luv\ and \alphaox\ are, by construction, correlated (B. Kelly
2005, personal communication).  It has been suggested that measuring
the redshift evolution of \lx , controlling for \luv , is a better
measure of the evolution of \alphaox\ since this relation does not
include the aforementioned \alphaox\ - \luv\ correlation.  However, it
is not clear that the redshift evolution of \lx\ and \alphaox\ are
necessarily related, even when controlling for the effects of \luv\
(e.g., M. Akritas 2005, personal communication).  We examine the redshift
evolution of \lx\ using our sample and find a highly significant
partial correlation between \lx\ - $z$, controlling for \luv\
($\tau_{12,3} = 0.21, \sigma=8.7$).  If the \lx\ - $z$
partial correlation is indeed a proxy for \alphaox\ - $z$, then this
result conflicts with our finding in \S3.5 that \alphaox\ does not
evolve with redshift.  We examine this apparent paradox further using
both Monte Carlo simulations and a subset of our full data sample that
better covers the \luv\ - $z$ plane.

To determine if the \lx\ - $z$ correlation is real or fake, we ran a
set of simulations designed to assess the strength of partial
correlations expected for samples with similar \luv\ - $z$
distributions. We started by randomly assigning an \luv\ value in the
observed range ($29.0<$\luv $<32.5$) to each AGN in our sample.  We
then computed \lx\ for each source assuming the ILS bisector relation
in Equation~\ref{eqn_luv_lx_3}, including a random scatter comparable
to that observed.  We compute the \lx\ - $z$ partial correlation
(controlling for \luv ) for ten random samples to obtain an average
value for the strength and significance of the partial correlation. In
these realizations there is no \luv\ - $z$ dependence, by
construction.  We find average correlation strength of $\tau_{12,3} =
0.06 \pm 0.02$ with $\sigma = 2.2\pm0.7$, significantly lower than for
our actual sample. However, if we instead constrain the random \luv\ to follow
the observed \luv\ - $z$ dependence, the partial tau method finds a
more significant, \emph{but false}, partial correlation of
$\tau_{12,3} = 0.17 \pm 0.01$ with $\sigma = 7.0\pm0.5$.  These
simulations suggest that the strength of the correlation between \lx\
and redshift is an artifact of the strong \luv\ - $z$ relation
in flux-limited samples.  While the partial-correlation methods of
Kendall's tau are designed to remove the contribution of the third
variable (in this case \luv ) from the measurement of the correlation
between the first and second variables (\lx\ and $z$), when the
correlations between the controlling variable and the first and/or
second variable are very strong, apparently significant, but false, partial
correlations can arise.

To test further the \lx\ - $z$ relation, we limit our analysis to
sources with $29.3 < $\luv $<31$ and $z<3$, consisting of 187 sources
(see the subsample box in Figure~\ref{lum}).  This subsample more
completely fills the \luv\ - $z$ plane and thus is less affected by
the strong \luv\ - $z$ correlation present in flux-limited samples.
Using this subsample, we find \lx\ - $z$ partial-correlation results
of $\tau_{12,3} = 0.11$ with $\sigma = 2.8$, again showing that the
strength of the relation diminishes greatly when the flux-limited
nature of the sample is reduced or eliminated.  For comparison, the
\alphaox\ - \luv\ partial correlation decreases, as one would expect
from the smaller sample size and reduced \luv\ and $z$ range, but it
remains significant ($\tau_{12,3} = -0.29$, $\sigma = 7.0$).  From
these tests it is apparent that the redshift evolution of \lx\ is
strongly influenced by the strong correlation between \luv\ and $z$
and thus should not be used as a substitute for the \alphaox\ - $z$
relation.

\boldmath
\subsection{Investigating non-linear \alphaox\ correlations}
\unboldmath

From the previous three sections it is clear that \alphaox\ is most
significantly correlated with \luv , but does the slope of this
correlation remain the same over the entire range of UV luminosities,
or is the \alphaox - \luv\ relation non-linear?  From
Figure~\ref{uv_vs_x} it appears possible that the \lx - \luv\ slope
may become steeper at lower luminosities.  It is difficult to tell
from the plot, however, as lower-luminosity sources typically have
larger measurement errors.  Non-linearity may also be hinted at in the
residual plot in Figure~\ref{luv_vs_aox}.  While all of the $3\sigma$
error bars for the mean residuals in each luminosity bin are
consistent with zero, there is a slight curve seen in the mean
residuals, peaking around $10^{30}$ erg~s$^{-1}$~Hz$^{-1}$.

We investigate this potential change in the slope of the \alphaox -
\luv\ relation by dividing the sample in half using the median UV
luminosity ($\mbox{log}($\luv$) = 30.352$) and calculating the slope
of the best-fit line for both the high-luminosity and low-luminosity
halves of the sample.  We find the coefficient for the low-luminosity
half to be flatter ($-0.098 \pm 0.019$) than for the high-luminosity
half ($-0.152 \pm 0.019$).  We use Student's t-test to determine if
these two slopes are drawn from the same distribution.  We find only a
$5\%$ probability that the best-fit slopes for the high and
low-luminosity halves of the distribution are drawn from the same
parent distribution, suggesting that the slope of the \alphaox - \luv\
relation may be \luv\ dependent.  

We use this same method to determine if the \alphaox - \luv\ slope is
dependent on redshift.  We divide the full sample into low-redshift
($z \le 2$) and high-redshift ($z > 2$) subsamples, containing 230 and
103 sources, respectively.  We do not find a significant difference in
the best-fit slopes for the low-redshift ($-0.125\pm0.012$) and
high-redshift ($-0.147\pm0.020$) subsamples.  Using Student's t-test,
we find a $37\%$ chance that the slopes of the low-redshift and
high-redshift subsamples are drawn from the same parent distribution.

\newpage
\section{Discussion and Conclusions}

\subsection{Summary of AGN Sample}

In this paper, we examine the correlations between \luv , \lx ,
\alphaox , and redshift for optically-selected AGNs.  We extend the
coverage of the luminosity-redshift plane relative to the S05 sample
by adding 52 lower-luminosity AGNs discovered by the \mbox{COMBO-17}
survey, 46 high-luminosity, low-redshift ($z<0.5$) AGNs from the BQS,
and 19 high-redshift ($z>4$) AGNs from recently published, targeted
\xray /optical AGN studies.  Radio-loud AGNs and BALQSOs are excluded
from our sample whenever possible, and the effects of host-galaxy
contamination are largely removed for the \mbox{COMBO-17} AGNs using
high-resolution ACS observations.  These additional AGNs bring the
total sample to 333 optically-selected AGNs, 293 ($88\%$) of which
have detected \xray\ counterparts.

In Table~\ref{sample_statistics} we present the statistical properties
of the full sample, broken into $\Delta \mbox{log(\luv )} = 1$ bins.
The Kaplan-Meier estimator within ASURV was used to calculate the
statistical properties of the sources in each bin.  The redshift
range, mean, RMS error, median, and both 25th and 75th percentile
values are given, along with the value of \luv\ and \alphaox\
calculated from Equations~(\ref{eqn_luv_lx_3}) and
(\ref{eqn_aox_luv}), respectively, using the mean \luv\ value given in
the appropriate bin.  This table also gives both the expected value of
\lx\ and \alphaox\ for each $\Delta \mbox{log(\luv )} = 1$ bin as well
as the spread around the expectation values.  From the table, it
appears that the spread of \lx\ decreases with increasing \luv\ for
the bins with statistically significant numbers of sources.  There is
no clear \luv\ dependence on the spread of \alphaox .  This table is
useful in identifying AGNs that may be \xray\ weak or assessing the
excess associated with jet-linked \xray\ emission.

\subsection{Results of Partial Correlation Analyses}

To measure correlations in our data we employ partial-correlation
methods that deal with censored data.  We confirm that the
monochromatic 2500~\AA\ and 2~keV luminosities are highly correlated
($15.3\sigma$), considering the effects of redshift, and follow the
relation $l_{\mbox{\scriptsize 2~keV}} \propto l_{\mbox{\scriptsize
    2500~\AA}}^{\beta}$, where $\beta = 0.73\pm0.01$ (from the ILS
bisector), and not $\beta = 1$ as found in some previous studies.

We investigate how the ratio of the \xray\ and UV luminosities,
\alphaox , is correlated with \luv , \lx , and redshift for our AGN
sample.  When taking into account the effects of redshift, we confirm
that the \alphaox - \luv\ anti-correlation is highly significant
($13.6\sigma$) but find a much smaller, but still significant,
anti-correlation for the \alphaox - \lx\ relation ($3.0\sigma$).  The
discrepancy in the significance of these two correlations is due, in
part, to the slope of the \lx - \luv\ relation being less than unity.
As can be seen in Figure~\ref{uv_vs_x}, the range of possible values
of \lx\ for a given value of \luv\ is smaller (and hence the range of
\alphaox\ values is smaller) than the range of possible values of
\luv\ for a given value of \lx .  In addition, some AGNs are more
variable in the \xray\ than they are in the optical/UV, with
long-timescale \xray\ observations of Seyfert~1s showing variability
$\ge 100\%$ for some sources \citep[][and references
therein]{uttley04}, which increases the possible ranges of \lx\ even
more dramatically for a given value of \luv .

Our analysis did not reveal any significant ($1.3\sigma$) evolution of
\alphaox\ with redshift (when considering the effects of \luv ), which
is consistent with the majority of previous \alphaox\ studies.
However, including the \mbox{COMBO-17} AGNs in our study allows us to
constrain, for the first time, the spectral evolution of
moderate-luminosity AGNs out to high redshifts.  Since
moderate-luminosity AGNs are the numerically dominant type of AGN in
the Universe, and \xray\ surveys have found luminosity-dependent
density evolution for such AGNs, it is important to understand the
spectral evolution of these types of sources.  The lack of cosmic
evolution of \alphaox\ also suggests that the energy generation
mechanisms that create AGN emission locally are already in place at $z
\simeq 5-6$.  This agrees with AGN studies which find no significant
evolution in the continuum shape of AGNs at high-redshifts from radio
\citep[e.g.,][]{petric03}, optical/UV \citep[e.g.,][]{pentericci03},
and \xray\ \citep[e.g.,][]{page05b,shemmer05a} studies.

\boldmath
\subsection{Comparison with X-ray--Selected, $z>4$ AGNs}
\unboldmath

\begin{deluxetable}{lccc}
  \tablecolumns{4}
  \tablecaption{\label{correlations}Partial Correlation Results.}
  \tablehead{ \colhead{Relation} & \colhead{Controlling} &
  \colhead{Kendall's} & \colhead{Significance} \\ \colhead{} &
  \colhead{Variable} & \colhead{$\tau_{12,3}$} & \colhead{} }
  \startdata
\cutinhead{Total Sample}
  \lx\ vs. \luv & $z$ & \phs0.519 & $15.3\sigma$ \\
  \alphaox\ vs. \luv & $z$ & $-0.377$ & $13.6\sigma$  \\
  \alphaox\ vs. \lx  & $z$ & $-0.080$ & \phn$3.0\sigma$ \\
  \smallskip \alphaox\ vs. $z$ & \luv & $-0.031$ & \phn$1.3\sigma$ \\
\cutinhead{Removed COMBO-17 AGNs with $\Gamma < 1.6$}
  \lx\ vs. \luv & $z$ & \phs0.505 & $13.7\sigma$ \\
  \alphaox\ vs. \luv & $z$ & $-0.396$ & $13.7\sigma$  \\
  \alphaox\ vs. \lx  & $z$ & $-0.079$ & \phn$3.0\sigma$ \\
  \smallskip \alphaox\ vs. $z$ & \luv & $-0.004$ & \phn$0.2\sigma$ \\
  \enddata
\tablecomments{Kendall's partial-rank correlation is defined as
  $\tau_{12,3} = \frac{\tau_{12}-\tau_{13}\tau_{23}}{\left[ \left(
        1-\tau_{13}^{2} \right) \left(1-\tau_{23}^{2} \right) \right]^{1\!/\!2}
  }$, where $\tau_{{\rm xy}}$ is the Kendall rank-correlation coefficient
  between data vectors $x$ and $y$ \citep{akritas96a}.}
\end{deluxetable}

%
%
\begin{figure}
\epsscale{1.1}
\plotone{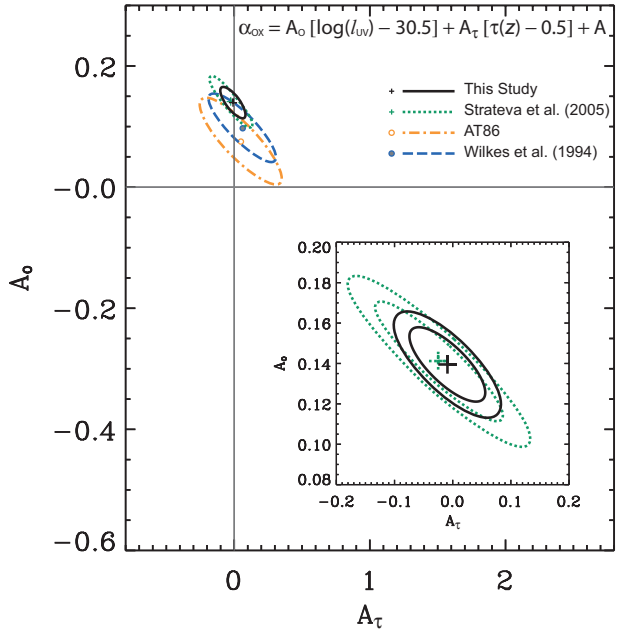}
\caption{\label{at_contours} Best-fit values and $\Delta S = 4.61$
  ($90\%$) confidence contours for the coefficients $A_{\mbox{\tiny
      O}}$ and $A_\tau$ for our sample ({\slshape black cross, solid
    black contour}), the sample of S05 ({\slshape green
    cross, dotted green contour}), and the samples examined by
  \citet{avni86} ({\slshape open orange circle, dot-dashed orange
    contour}) and \citet{wilkes94} ({\slshape filled blue circle,
    dashed blue contour}).  {\slshape Inset:} A magnified view of the
  contours for our sample.  Both the $68\%$ ($\Delta S=2.30$) and
  $90\%$ ($\Delta S=4.61$) confidence contours ({\slshape solid black
    contours}) are shown.  For comparison, the $68\%$ and $90\%$
  confidence contours of the S05 sample are also shown
  ({\slshape dotted green contours}).}
\end{figure}

In \S3, we limited our analyses and discussion to the 333
optically-selected AGNs in our sample.  Here we determine if the
relations we derived for \luv , \lx , \alphaox , and redshift also
hold for the eight published \xray --selected, radio-quiet, $z>4$ AGNs
(see \S2.2.3).  While subject to different selection biases than
our main AGN sample, these \xray--selected AGNs substantially extend
the coverage of the luminosity - redshift plane (see {\slshape
  crosses} in Figure~\ref{lum}a), and thus they can be useful in
determining if the aforementioned relations are also valid for
high-redshift, low-luminosity AGNs.

From Figures~\ref{uv_vs_x}, \ref{luv_vs_aox}, and \ref{lx_vs_aox} it
is apparent that the \xray--selected AGNs, while considerably fainter
in \luv\ than the optically-selected AGNs in our sample at similar
redshifts, have \alphaox\ values that do not significantly stand out
from the \alphaox\ values of the optically-selected AGNs.  While the
\xray --selected AGNs do populate the high (less-negative) portion of
the \alphaox\ distributions in Figures~\ref{luv_vs_aox} and
\ref{lx_vs_aox}, this is not unexpected because \xray\ selection is
biased toward sources that are \xray\ bright.  In other words, for a
group of sources at a given \luv\ with the observed spread in \alphaox
, \xray\ surveys are biased toward selecting those sources with
higher \alphaox\ values (i.e., the AGNs that are relatively brighter in
\xray s).

In Figure~\ref{z_vs_aox}, the \xray--selected sources all lie well above
the best-fit \alphaox\ - $z$ relation derived from our main sample.
This apparent discrepancy is {\slshape expected} if \alphaox\ is
indeed dependent on \luv\ and not $z$.  The eight \xray--selected AGNs
have considerably smaller \luv\ values than the optically selected,
$z>4$ AGNs, and thus we would predict that they would have higher
(less negative) values of \alphaox\ from Equation~(\ref{eqn_aox_luv}).
Indeed, the \alphaox\ residuals shown in Figures~\ref{residuals} and
\ref{residuals_lum_breakdown} demonstrate that the \xray--selected
AGNs ({\slshape crosses}) agree fairly well with our conclusion that
\alphaox\ is dependent on \luv\ and shows no significant redshift
dependence.  To test further this conclusion we again ran the partial
correlations described in \S3, this time including the eight
\xray--selected sources.  We find that the significance of most of
the correlations presented in Table~\ref{correlations} increase
$\sim0.3-0.8\sigma$.  The \alphaox\ - $z$ partial correlation
increases by $\sim0.7\sigma$, but it is still below the threshold of a
significant correlation. In addition, while the \xray--selected AGNs
do increase the coverage of the luminosity - redshift plane, it is not
clear what effect the different selection biases of the
\xray--selected AGNs will have on these correlations.

\begin{deluxetable*}{lcccccc}
  \tabletypesize{\scriptsize}
  \tablecolumns{7}
  \tablecaption{\label{sample_statistics}Sample Statistics}
  \tablehead{ \colhead{log(\luv) Range} & \colhead{$27-28$} &
  \colhead{$28-29$} & \colhead{$29-30$} & \colhead{$30-31$} &
  \colhead{$31-32$} & \colhead{$32-33$} } 
\startdata 
Number of Sources & 6 & 41 & 84 & 121 & 67 & 14 \\ 
Number of \xray\ Detections & 5 & 38 & 82 & 99 & 55 & 14 \\
\xray\ Detection Fraction  & \phn$83\%$ & \phn$93\%$ & \phn$98\%$ & \phn$82\%$ &\phn $82\%$ & $100\%$ \\
Redshift Range & $0.009-0.201$ & $0.009-1.264$ & $0.033 - 2.870$ & $0.085 - 4.190$ & $0.472 - 6.280$ & $2.736 - 4.550$ \\
\cutinhead{log(\luv)}
Mean log(\luv)       &     27.847 &    28.604 &    29.622 &    30.540 &    31.552 &    32.259 \\
RMS log(\luv)        &  \phn0.081 & \phn0.250 & \phn0.238 & \phn0.286 & \phn0.292 & \phn0.150 \\
25th Percentile      &     27.753 &    28.508 &    29.453 &    30.283 &    31.264 &    32.120 \\
Median log(\luv)     &     27.825 &    28.630 &    29.625 &    30.609 &    31.523 &    32.200 \\
75th Percentile      &     27.890 &    28.781 &    29.806 &    30.791 &    31.742 &    32.330 \\

\cutinhead{log(\lx)}
Mean log(\lx)        & 24.571 & 25.149 & 25.953 & 26.472\tablenotemark{a} & 27.105\tablenotemark{a} & 27.600 \\
RMS log(\lx)         &  \phn0.318 & \phn0.621 & \phn0.431 & \phn0.418 & \phn0.333 & \phn0.348 \\
25th Percentile & \nodata \tablenotemark{b} & 24.873 & 25.792 & 26.296  & 26.879  & 27.340 \\
Median log(\lx) & 24.463 & 25.222 & 26.018 & 26.462  & 27.065  & 27.730 \\
75th Percentile & 24.661 & 25.622 & 26.221 & 26.746 & 27.337 & 27.836 \\
Value from Equation~(\ref{eqn_luv_lx_3}) & 24.749 & 25.235 & 25.889 & 26.478 & 27.128 & 27.582 \\

\cutinhead{\alphaox}
Mean \alphaox   & $-1.253$ & $-1.322$ & $-1.408$ & $-1.568$\tablenotemark{a} & $-1.708$ & $-1.788$ \\
RMS \alphaox    & \phs0.093 & \phs0.192 & \phs0.165 & \phs0.198 & \phs0.146 & \phs0.131 \\
25th Percentile         & $-1.343$ & $-1.451$ & $-1.493$ &  $-1.657$ & $-1.810$ & $-1.865$ \\
Median \alphaox         & $-1.309$ & $-1.291$ & $-1.402$ &  $-1.547$ & $-1.691$ & $-1.786$ \\
75th Percentile         & $-1.230$ & $-1.184$ & $-1.289$ &  $-1.455$ & $-1.609$ & $-1.717$ \\
Value from Equation~(\ref{eqn_aox_luv}) & $-1.189$ & $-1.293$ & $-1.433$ &  $-1.559$ & $-1.699$ & $-1.796$ \\
  \enddata
\tablenotetext{a}{ASURV changed first datum in bin from a censored point to a detection --- mean estimate is biased.}
\tablenotetext{b}{No value is reported due to bias by the upper limit and the small number of sources in this bin.}
\end{deluxetable*}

\subsection{Comparison with Earlier Studies}

The AGN sample presented here enables us to provide the tightest
constraints to date on the contributions of \luv\ and redshift to
\alphaox .  To illustrate our improvement over some past studies, we
compare our results with those of \citet[hereafter AT86]{avni86},
\citet{wilkes94}, and S05.  Note that the first two
authors use cosmological look-back time, $\tau(z)$, in units of the
present age of the Universe, and not redshift in their calculations.
We calculate $\tau(z)$ for our sources and use the method outlined in
\S\S~3~and~4 of AT86 to calculate confidence contours.  We rename
$A_{\mbox{\tiny Z}}$ from AT86 to $A_{\tau}$ since, strictly speaking,
it is the coefficient of cosmological look-back time which is not a
linear function of redshift.  This modification changes
Equation~(\ref{eqn_aox_luv_tau}) of AT86\footnotemark\ to
\footnotetext{In AT86, \alphaox\ is defined to be positive. Thus, the
  coefficients $A_{\mbox{\tiny O}}$ and $A_{\tau}$ have the opposite
  sign as the coefficients presented the previous equations.}
\begin{eqnarray}
\lefteqn{\overline{\alpha}_{\mbox{\tiny OX}}(l_{\mbox{\scriptsize 2500 \AA}},z \mid  \mbox{X-ray Loud}) = } \nonumber \\
 & & A_{\mbox{\tiny O}}\left[\mbox{log}\left(l_{\mbox{\scriptsize 2500 \AA}}\right) - 30.5 \right] + A_{\tau}\left[ \tau \left(z\right) - 0.5 \right] + A \label{eqn_aox_luv_tau}
\end{eqnarray}

For our sample, we find best-fit values of $[A_{\mbox{\tiny
    O}},A_{\tau},A] =
[0.140^{+0.011}_{-0.016},-0.009^{+0.055}_{-0.040},1.553\pm{0.014}]$
for Equation~\ref{eqn_aox_luv_tau}.  In Figure~\ref{at_contours}, we
show the $\Delta S=4.61$ confidence contours for our data and the
samples presented in AT86, \citet{wilkes94}, and S05.  These are
equivalent to $90\%$ confidence contours taking two parameters to be
of interest.  The tighter constraints provided by our data have a
number of sources.  Our \xray\ detection fraction ($88\%$) is higher
than either AT86 ($61\%$) or \citet{wilkes94} ($64\%$).  Unlike early
AGN samples used to study \alphaox , our data represent a relatively
homogeneous collection of AGNs where we attempted to remove or reduce
sources of systematic errors (radio jets, absorption, galaxy
contribution, etc.).  Finally, to our knowledge, our AGN sample
provides the best coverage of the luminosity - redshift plane to date.

\subsection{Comparing \xray\ and Optical AGN Luminosity Functions}

Understanding the relationship between \lx\ and \luv\ in AGNs aids in
explaining discrepancies between the AGN X-ray and optical luminosity
functions (XLF and OLF, respectively).  For luminous AGNs, both the
OLF \citep[e.g.,][]{boyle00,croom04,richards05,richards06} and XLF
\citep[e.g.,][]{cowie03,ueda03,barger05a,hasinger05,lafranca05} are
consistent with pure luminosity evolution (PLE), with the peak
activity of luminous AGNs occurring at $z=2-3$.  However, recent X-ray
surveys have found that low-luminosity ($L_{2-8 {\rm ~keV}} \simlt
10^{44}$ erg~s$^{-1}$) AGNs do not agree with the PLE model, but
instead appear to undergo luminosity-dependent density evolution
\citep[LDDE;
e.g.,][]{ueda03,steffen03,barger05a,hasinger05,lafranca05,silverman05b}.
These studies found that the peak density of low-luminosity AGNs
occurs at lower redshifts than for high-luminosity AGNs, an example of
``cosmic-downsizing''.  LDDE is not seen in deep optical AGN surveys
where the OLFs of low-luminosity AGNs are not well constrained and are
consistent with both PLE and pure density evolution (PDE) models
\citep{wolf03b}.
 
Given that AGNs found in deep \xray -selected samples appear to
undergo LDDE, does this necessitate LDDE for optically-selected
samples as well?  We can use the LDDE equations given in \S~5.2 of
\citet[][hereafter U03]{ueda03} and the relationship between \lx\ and
\luv\ we find for optically-selected AGNs
(Equation~\ref{eqn_luv_lx_3}) to examine how the LDDE parameters
change if the optical luminosity is substituted for the X-ray
luminosity.  Equations~(16) and (17) of U03 give the evolution factor,
$e(L_X,z)$, as a function of both the luminosity and redshift, which
is unique to LDDE models (for PLE and PDE, the evolution factor only
depends upon redshift).  More specifically, the redshift cutoff,
$z_c$, included in the evolution factor is a function of luminosity in
LDDE models and is a constant in PLE and PDE models.  Applying our \lx
- \luv\ relation to the redshift cutoff (Equation 17 of U03), we find
that the power-law index, $\alpha$, becomes $\alpha^\prime = \alpha
\cdot \beta$, where $\beta$ is the slope of the \lx - \luv\ relation,
when the X-ray luminosity is replaced with the optical/UV luminosity.
The evolution factor becomes independent of optical/UV luminosity
[i.e., $e(L_{UV},z) \to e(z)$] only when the \xray\ and optical/UV
emission in AGNs are independent of one another (i.e., $\beta = 0$).
Thus, aside from this special case, the LDDE observed for \xray\ AGN
samples implies LDDE for optical AGN samples.

The $\beta = 0.725$ slope found for the \luv - \lx\ relation
(Equation~\ref{eqn_luv_lx_3}) evinces that a group of AGNs will span a
larger range in UV luminosity than in \xray\ luminosity.  In other
words, optical/UV surveys must cover a larger luminosity range than
X-ray surveys to probe the same AGN populations.  This is apparent
when comparing the sky densities of AGNs from deep \xray\ and optical
surveys (see \S~1).  The deep X-ray surveys find about an order of
magnitude more AGNs per square degree than the deepest optical
surveys.  Some of this discrepancy is attributable to different AGN
selection effects between the X-ray and optical surveys.  Host-galaxy
contamination is much more prevalent in the optical regime, as is
intrinsic obscuration.  However, if we compare the depth probed by the
optical and X-ray surveys we see that the optical surveys are missing
the low-luminosity AGNs that do not fit into the PLE paradigm.
Studies of the XLF find the densities of AGNs with $L_{2-8 {\rm ~keV}}
= 10^{42} - 10^{44} \mbox{ erg s}^{-1}$ peak at lower redshifts than
their more luminous cousins, a strong indication of LDDE.  This range
in hard \xray\ luminosity corresponds to a monochromatic \lx\ range of
log(\lx )$ = 23.9 - 25.9$, assuming $\Gamma = 2$.  Using
Equation~(\ref{eqn_luv_lx_3}), we find the corresponding monochromatic
\luv\ range to be log(\luv)$ = 26.9 - 29.7$.  From Figure~\ref{lum}a
it is apparent that the optically-selected AGN samples do not extend
to faint enough UV luminosities to detect all of these AGNs.  Even if
we assume $\beta=1$, we find that the optical surveys are still too
shallow, but the problem becomes less severe.  Assuming $\beta=1$, the
range of \lx\ covered by the X-ray selected AGN surveys corresponds to
a monochromatic \luv\ range of log(\luv)$ = 27.8 - 29.8$, an order of
magnitude higher than we get with $\beta=0.73$. In light of this
comparison it is not surprising that LDDE is not required to describe
the evolution of the OLF, because optical surveys are not deep enough
to probe the AGN luminosities where LDDE becomes apparent in XLF
studies.

\subsection{Future Studies}
This study presents the highest significance \alphaox\ - \luv\
correlations to date, but significant work remains to be done.  The
significance of the measured correlations could be substantially
improved by eliminating the intrinsic scatter in our data due to UV
and \xray\ variability.  This could be accomplished with concurrent UV
and \xray\ observations of AGNs with, e.g., \xmm .  This study could
also be extended by including more of the most optically luminous
quasars known at $z\sim1-4$.  The high luminosities of these quasars
make them visible at all redshifts, and their inclusion would reduce
the paucity of these sources at low-to-moderate redshifts in our
current sample.  We could also extend our study to include Low
Ionization Nuclear Emission-line Regions (LINERs).  These
low-luminosity AGNs could better constrain the \alphaox\ - \luv\
relation, and the additional \luv\ ``leverage'' at low luminosities would
help test the possible non-linear \alphaox\ - \luv\ relation suggested
by our data.  To measure the AGN emission in LINERs it is essential to
minimize the contamination from the host galaxy. To do this, the
high-resolution capabilities of \hst\ and \chandra\ are needed to
isolate the intrinsic AGN emission within LINERs.  Finally, there are
currently no compelling physical models that explain the origin of the
observed \alphaox - \luv\ relation.  Any theoretical model attempting
to describe the emission mechanisms within AGNs must be able to
reproduce this observed relation.  There is hope that attempts to
model the environments around accreting SMBHs from first principles
using relativistic magneto-hydrodynamic simulations
\citep[e.g.,][]{hawley95,de_villiers03,krolik05} will be able to
explain the observed \alphaox\ - \luv\ relation, but to our knowledge
these simulations have not yet found a physical explanation for the
observed \alphaox\ - \luv\ relation.

\acknowledgements 
We thank the anonymous referee for helpful comments that improved the
manuscript.  We thank Wolfgang Voges and Thomas Boller for supplying
the $0.5-2$ keV \rosat\ count rates for the RASS BQS sample along with
Michael Akritas, Ohad Shemmer, and Brandon Kelly for helpful discussions.  We
gratefully acknowledge support from NSF CAREER award AST-9983783
(A.~T.~S. and W.~N.~B.), CXC grant GO4-5157A (A.~T.~S., W.~N.~B.,
B.~D.~L., and D.~P.~S.), NASA LTSA grant NAG5-13035 (I.~S. and
W.~N.~B.), the Royal Society (D.~M.~A), and MIUR COFIN grant 03-02-23
(C.~V.)

Funding for the creation and distribution of the SDSS Archive has been
provided by the Alfred P. Sloan Foundation, the Participating
Institutions, the National Aeronautics and Space Administration, the
National Science Foundation, the U.S. Department of Energy, the
Japanese Monbukagakusho, and the Max Planck Society. The SDSS Web site
is {\ttfamily http://www.sdss.org/}.

The SDSS is managed by the Astrophysical Research Consortium (ARC) for
the Participating Institutions. The Participating Institutions are The
University of Chicago, Fermilab, the Institute for Advanced Study, the
Japan Participation Group, The Johns Hopkins University, the Korean
Scientist Group, Los Alamos National Laboratory, the
Max-Planck-Institute for Astronomy (MPIA), the Max-Planck-Institute
for Astrophysics (MPA), New Mexico State University, University of
Pittsburgh, University of Portsmouth, Princeton University, the United
States Naval Observatory, and the University of Washington.

{\it Facilities:} \facility{CXO (ACIS)}, \facility{HST (ACS)},
\facility{Max Plank:2.2m (WFI)}, \facility{VLA}, \facility{ROSAT
  (PSPC)}, \facility{Sloan}

\bibliographystyle{aj}
\bibliography{ms}

\begin{appendix}
\boldmath
\section{Properties of Additional \lowercase{$z>4$} AGNs}
\unboldmath

We list the properties of the publicly available sources from
\citet{kelly05} in Table~\ref{tab1}.  To our knowledge, details on
these sources are not available in the literature.  The \chandra\ data
were reduced using the same methods outlined in \citet{vignali01}.  In
Table~\ref{xrs}, we list the properties of the \xray --selected AGN
presented in \S~2.2.3.  

\suppressfloats[t]

\tabletypesize{\scriptsize}
\setlength{\tabcolsep}{2pt}
\begin{deluxetable*}{lcrcccccccccclr}[ht]
\setlength{\tabcolsep}{3pt}
\tablecolumns{15}
\tabletypesize{\scriptsize}
\tablecaption{Properties of $z>4$ Quasars Observed by \chandra}
\tablehead{ 
 \colhead{Object} & 
 \colhead{$z$} &
 \colhead{$N_{\rm H}$} & 
 \colhead{$AB_{1450}$} & 
 \colhead{$f_{2500}$} & 
 \colhead{$\log (L_{2500})$} & 
 \colhead{$M_B$} & 
 \colhead{Exp.~Time} &
 \colhead{Count~rate} & 
 \colhead{$f_{\rm SB}$} & 
 \colhead{$f_{\rm 2\/keV}$} & 
 \colhead{$\log (\nu L_\nu )_{\rm 2\/keV}$} & 
 \colhead{$\log (L_{\rm HB})$} & 
 \colhead{$\alpha_{\rm ox}$} & 
 \colhead{$R$} \\
 \colhead{(1)} & 
 \colhead{(2)} & 
 \colhead{(3)} & 
 \colhead{(4)} & 
 \colhead{(5)} & 
 \colhead{(6)} & 
 \colhead{(7)} &  
 \colhead{(8)} & 
 \colhead{(9)} & 
 \colhead{(10)} & 
 \colhead{(11)} & 
 \colhead{(12)} & 
 \colhead{(13)} &
 \colhead{(14)} &
 \colhead{(15)}
}
\startdata
QSO~0910$+$564  & 4.04    & 2.73 & 20.7 & 3.08 & 30.98 & $-$25.9 & 22849 & 0.34$^{+0.17}_{-0.12}$ & 1.32$^{+0.67}_{-0.45}$ & 0.99 & 44.18 & 44.4 & $-1.72\pm{0.09}$        & $<16.6$ \\
PC~1450$+$3404  & 4.19    & 1.26 & 21.0 & 2.33 & 30.89 & $-$25.7 & 14842 & 0.94$^{+0.32}_{-0.25}$ & 3.49$^{+1.21}_{-0.93}$ & 2.71 & 44.64 & 44.9 & $-1.51^{+0.08}_{-0.07}$ & $<20.3$ \\
SDSS~1413$+$0000 & 4.08   & 3.12 & 19.8 & 7.05 & 31.35 & $-$26.8 & 11843 & 1.17$^{+0.40}_{-0.32}$ & 4.61$^{+1.58}_{-1.25}$ & 3.49 & 44.74 & 44.9 & $-1.65\pm{0.07}$        & $<7.4$  \\
%
%
SDSS~0050$-$0053 & 4.33 & 2.69 & 19.5 & 9.73 & 31.53 & $-$27.3 & 12737 & 1.15$^{+0.39}_{-0.29}$ & 4.24$^{+1.44}_{-1.08}$ & 3.38 & 44.75 & 44.9 & $-1.71\pm{0.07}$ & $<5.6$ \\
SDSS~1444$-$0123 & 4.18 & 3.92 & 19.5 & 9.64 & 31.50 & $-$27.2 & 10001 & 0.40$^{+0.32}_{-0.19}$ & 1.60$^{+1.28}_{-0.76}$ & 1.24 & 44.30 & 44.5 & $-1.87^{+0.11}_{-0.12}$ & $<5.4$ \\
SDSS~2357$+$0043 & 4.36 & 3.28 & 19.8 & 7.38 & 31.41 & $-$27.0 & 12657 & 0.85$^{+0.35}_{-0.25}$ & 3.47$^{+1.43}_{-1.03}$ & 2.73 & 44.67 & 44.9 & $-1.70\pm{0.08}$        & $<7.0$  \\
\tableline
\enddata
\tablecomments{
(1) Source name; 
(2) Redshift;
(3) Galactic column density, from Dickey \& Lockman (1990), in units of $10^{20}$~cm$^{-2}$; 
(4) AB magnitude at rest-frame $1450$~\AA; 
(5) flux density at rest-frame 2500~\AA, in units of $10^{-28}$~erg~cm$^{-2}$~s$^{-1}$~Hz$^{-1}$; 
(6) log of the monochromatic luminosity at rest-frame 2500~\AA, in units of erg~s$^{-1}$~Hz$^{-1}$; 
(7) absolute $B-$band magnitude; 
(8) Chandra exposure times (in seconds) corrected for detector dead time and high-background periods;
(9) observed count rate computed in the soft band, in units of $10^{-3}$ counts~s$^{-1}$; 
(10) Galactic absorption-corrected flux in the observed 0.5--2 keV band,  in units of $10^{-15}$~erg~cm$^{-2}$~s$^{-1}$; 
(11) rest-frame 2~keV flux density, in units of $10^{-32}$~erg~cm$^{-2}$~s$^{-1}$~Hz$^{-1}$; 
(12) log of the luminosity at rest-frame 2~keV, in units of erg~s$^{-1}$; 
(13) log of the 2--10~keV rest-frame luminosity, corrected for the effect of Galactic absorption, in units of erg~s$^{-1}$; 
(14) optical-to-X-ray spectral index; errors have been computed following the ``numerical method'' described in $\S$~1.7.3 
of Lyons (1991); both the statistical uncertainties on the \xray\ count rates and the effects of the observed ranges of the 
\xray\ and optical continuum shapes have been taken into account; 
(15) radio loudness parameter, defined as $R$ = $f_{\rm 5~GHz}/f_{\rm 4400~\mbox{\scriptsize\AA}}$ (rest frame). 
The 5~GHz flux density is computed from the 1.4~GHz flux density (from FIRST) 
assuming a radio power-law slope of $\alpha=-0.8$, with $f_{\nu}\propto~\nu^{\alpha}$; 
}
\label{tab1}
\end{deluxetable*}

\setlength{\tabcolsep}{4pt}
\begin{deluxetable*}{lcccccccccl}[ht]
\tablecolumns{11}
\tabletypesize{\scriptsize}
\tablecaption{Properties of \xray--Selected, $z>4$ AGNs}
\tablehead{ 
 \colhead{Object} & 
 \colhead{$z$} &
 \colhead{$N_{\rm H}$} & 
 \colhead{$AB_{1450}$} & 
 \colhead{$f_{2500}$} & 
 \colhead{$\log (L_{2500})$} & 
 \colhead{$f_{\rm SB}$} & 
 \colhead{$f_{\rm 2\/keV}$} & 
 \colhead{$\log (L_{\rm 2\/keV})$} & 
 \colhead{$\alpha_{\rm ox}$} &
 \colhead{Reference} \\
 \colhead{(1)} & 
 \colhead{(2)} & 
 \colhead{(3)} & 
 \colhead{(4)} & 
 \colhead{(5)} & 
 \colhead{(6)} & 
 \colhead{(7)} & 
 \colhead{(8)} & 
 \colhead{(9)} & 
 \colhead{(10)} &
 \colhead{(11)} 
}
\startdata
CXOCY J033716.7$-$050153   & 4.61 & 4.82 & 23.8 & $-$28.84  & 29.74  & 1.8\phn & $-$31.82 & 26.76 & $-$1.14 & \citet{treister04a} \\


CLASXS J1032414.2$+$572227 & 5.40 & 0.73 & 24.3 & $-$29.04  & 29.65  & 0.85 & $-$32.09 & 26.60 & $-$1.17 & \citet{steffen04,yang04} \\

RX J1052$+$5719            & 4.45 & 0.56 & 22.6 & $-$28.36  & 30.20  & 2.3\phn & $-$31.73 & 26.83 & $-$1.29 & \citet{schneider98} \\

CXOMP J105655.1$-$034322   & 4.05 & 3.55 & 22.3 & $-$28.24  & 30.26  & 5.2\phn & $-$31.37 & 27.13 & $-$1.20 & \citet{silverman05a} \\

CXOHDFN J123647.9$+$620941 & 5.19 & 1.48 & 23.5 & $-$28.72  & 29.94  & 0.29 & $-$32.57 & 26.09 & $-$1.48 & \citet{barger02,vignali02} \\

CXOHDFN J123719.0$+$621025 & 4.14 & 1.45 & 25.2 & $-$29.40  & 29.11  & 0.26 & $-$32.70 & 25.81 & $-$1.26 & \citet{barger02,vignali02} \\

CXOCY J125304.0$-$090737   & 4.18 & 2.96 & 23.0 & $-$28.52  & 30.00  & 1.7\phn & $-$31.88 & 26.64 & $-$1.28 & \citet{castander03} \\


CXOMP J213945.0$-$234655   & 4.93 & 3.55 & 21.6 & $-$27.96  & 30.67  & 1.8\phn & $-$31.80 & 26.83 & $-$1.47 & \citet{silverman02} \\

\tableline
\enddata
\tablecomments{
(1) Source name (in order of increasing Right Ascension); 
(2) Redshift;
(3) Galactic column density, from Dickey \& Lockman (1990), in units of $10^{20}$~cm$^{-2}$; 
(4) AB magnitude at rest-frame $1450$~\AA; 
(5) log of the flux density at rest-frame 2500~\AA, in units of erg~cm$^{-2}$~s$^{-1}$~Hz$^{-1}$; 
(6) log of the monochromatic luminosity at rest-frame 2500~\AA, in units of erg~s$^{-1}$~Hz$^{-1}$; 
(7) Galactic absorption-corrected flux in the observed 0.5--2 keV band,  in units of $10^{-15}$~erg~cm$^{-2}$~s$^{-1}$; 
(8) log of the rest-frame 2~keV flux density, in units of erg~cm$^{-2}$~s$^{-1}$~Hz$^{-1}$; 
(9) log of the monochromatic luminosity at rest-frame 2~keV, in units of erg~s$^{-1}$~Hz$^{-1}$; 
(10) optical-to-X-ray spectral index;
(11) References for \xray--selected AGN.
}
\label{xrs}
\end{deluxetable*}

\end{appendix}

%
\end{document}